\documentclass[11pt,letterpaper]{article}
\usepackage[margin=1in]{geometry}
\usepackage{times}
\usepackage{natbib}
\setcitestyle{authoryear,round,citesep={;},aysep={,},yysep={;}}
\usepackage[T1]{fontenc}
\usepackage{amsmath,amssymb,booktabs,array,multirow,graphicx,longtable,listings}
\DeclareMathSizes{8.3}{8.3}{6}{5}
\usepackage{xcolor}
\usepackage{xspace}
\usepackage{hyperref}
\usepackage{url}
\usepackage{xurl}
\usepackage{placeins}
\usepackage{float}
\hypersetup{colorlinks=true,linkcolor=black,citecolor=black,urlcolor=black,
  pdftitle={Fix What's Wrong, Keep What's Right: Hallucination Repair for LLM-Based Optimization Modeling},
  pdfauthor={Jiageng Wu, Chao Shen, Zihan Guo, Junyan Shao, Zongyan Zhang, Jie Song, Zhong Li, Mingyang Sun}}
\newcommand{\method}{\mbox{OPT-Kraken}\xspace}

\newcommand{\clean}{\textsc{Clean}\xspace}
\newcommand{\err}{\textsc{Error}\xspace}
\newcommand{\unres}{\textsc{Unresolved}\xspace}

\title{Fix What's Wrong, Keep What's Right: Hallucination Repair for LLM-Based Optimization Modeling}
\author{Jiageng Wu$^{1}$ \quad Chao Shen$^{2}$ \quad Zihan Guo$^{1}$ \quad Junyan Shao$^{1}$\\
Zongyan Zhang$^{1}$ \quad Jie Song$^{1}$ \quad Zhong Li$^{3,*}$ \quad Mingyang Sun$^{1,*}$\\[0.7em]
\small $^{1}$Peking University \quad $^{2}$Zhejiang University \quad $^{3}$Great Bay University\\[0.2em]
\small $^{*}$Corresponding authors: Zhong Li (\href{mailto:zhongli@gbu.edu.cn}{zhongli@gbu.edu.cn})\\
\small and Mingyang Sun (\href{mailto:smy@pku.edu.cn}{smy@pku.edu.cn})}
\date{}
\begin{document}
\maketitle
\begin{abstract}
Large language models (LLMs) are increasingly being explored to automate operations research modeling, reduce modeling time and lower the expertise barrier to optimization based decision support. However, this process often introduces inconsistencies between generated artifacts and the original problem, which we refer to as optimization modeling hallucinations. Despite progress in general-purpose hallucination mitigation, optimization modeling poses a distinct repair challenge: restoring fidelity to the original problem through coordinated model--code updates. We introduce \method, a training-free hallucination repair framework. To the best of our knowledge, it is the first post-hoc framework for repairing hallucinations in LLM-based optimization modeling artifacts. It extracts problem requirements and instance data, localizes errors by checking problem--model fidelity, model--code consistency, and data grounding, and selects repair granularity by the affected scope. For local repair, category rules and instance evidence are compiled into contracts specifying permitted edits, expected effects, and protected content. Experiments on established optimization modeling benchmarks show positive gains across automatic modeling frameworks and fine-tuned models. On zero-shot artifacts from multiple base models, \method effectively repairs optimization modeling hallucinations while preserving initially correct artifacts. These results demonstrate effective and controlled repair for LLM-based optimization modeling artifacts.
\end{abstract}

\section{Introduction}
Large language models (LLMs) are increasingly being explored to translate natural language operations research (OR) problems into mathematical models $M$ and solver code $S$. This workflow can lower the expertise barrier and reduce the modeling time required for decision support in manufacturing, logistics, and resource allocation. However, $M$ and $S$ generated by LLM often fail to remain faithful to the original problem, potentially misleading downstream decisions. We define \emph{optimization modeling hallucinations} as errors in LLM-generated modeling artifacts that make them inconsistent with the original optimization problem. These include unsupported additions, omitted requirements, incorrect mathematical formulations, model--code inconsistencies, data-binding faults, and result-extraction errors, among other failures.

General hallucination research provides foundations for evidence checking, consistency estimation, and feedback-driven revision. However, unlike question-answering tasks whose primary outputs are general natural language responses, optimization modeling requires translating problem requirements into interrelated mathematical models and solver code. Merely checking support for generated statements does not establish that all problem requirements have been represented; successful code execution cannot rule out errors in the mathematical model. Therefore, repairing optimization modeling hallucinations requires locating errors using evidence from the original problem and defining the scope of coordinated model--code updates.

\begin{figure}[t]
  \centering
  \includegraphics[width=\linewidth]{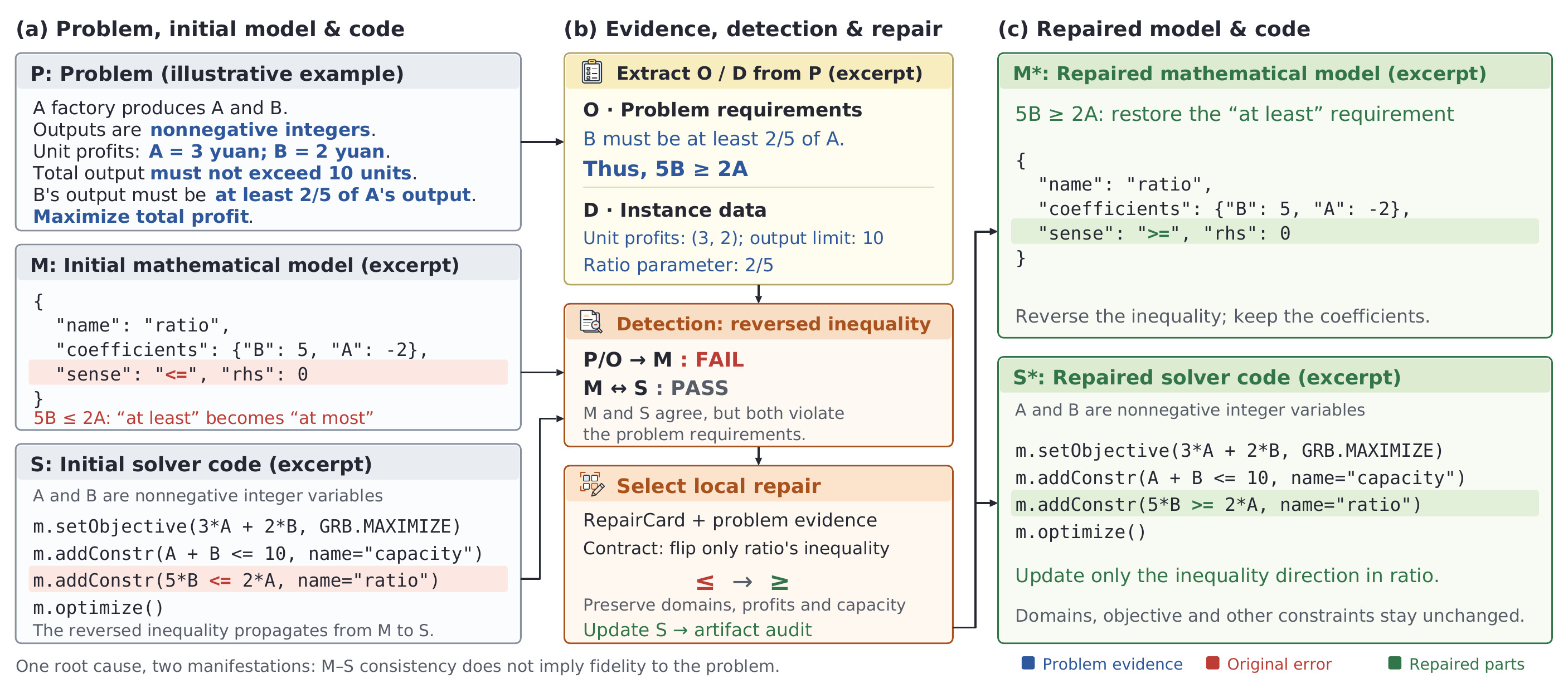}
  \caption{A shared optimization-modeling hallucination in the model and code.}
  \label{fig:example}
\end{figure}

These requirements motivate treating hallucination repair as a standalone post-hoc task on optimization modeling artifacts. Rather than simply generating another model and code, the task is to identify errors in existing artifacts using evidence from the original problem, perform necessary repairs, and use validation to decide whether to accept the changes. Figure~\ref{fig:example} illustrates a shared modeling hallucination and its localized repair: both the mathematical model and solver code use $5B\leq2A$, whereas the original problem requires $5B\geq2A$. The repair corrects the inequality direction in both artifacts while leaving variable domains, objective coefficients, and other constraints unchanged.

We study hallucination repair through the joint lens of error correction and content preservation. Incorrect outputs deemed hallucination-free lose repair opportunities; false alarms can corrupt correct outputs through unnecessary edits. Repair must correct errors while preserving valid content. We therefore use three metrics: \textbf{Rescue}, \textbf{Regression}, and \textbf{Error Coverage}. Rescue is the share of initially incorrect outputs corrected, Regression is the share of initially correct outputs made incorrect, and Error Coverage is the share of initially incorrect outputs retained as repair opportunities by the screening stage. Together, they capture gains, outcome-level harm, and coverage of opportunities.

\method provides \emph{plug-and-play post-processing} of a problem $P$, model $M_0$, and code $S_0$, without retraining or changing the upstream generator. It extracts requirements $O$ and instance data $D$ from problem $P$, localizes errors through directional fidelity checks, and selects the repair scope. Contracts bind category rules to case evidence; commit gates protect valid content.

To the best of our knowledge, \method is the first post-hoc framework for repairing hallucinations in LLM-generated optimization-modeling artifacts, targeting problem--model--code fidelity while controlling repair scope and regression. Our contributions can be summarized as follows:

\noindent\textbf{(1) Post-hoc formulation of hallucination repair for LLM-based optimization modeling:} We formulate hallucination repair for LLM-based optimization modeling as a post-hoc transformation of frozen problem--model--code artifacts. Unlike generation-time correction, the task requires correcting unfaithful content while preserving valid content. We evaluate repair effectiveness, preservation, and coverage of repair opportunities using Rescue, Regression, and Error Coverage, defined respectively as the shares of initially incorrect outputs corrected, initially correct outputs made incorrect, and initially incorrect outputs flagged by screening for examination or potential repair.

\noindent\textbf{(2) Evidence-guided and scope-controlled hallucination repair framework:} We introduce \method, a generic training-free framework that operates as a post-processor for different upstream generators. It (a) diagnoses artifacts through source-grounded checks of problem--model fidelity, model--code consistency, and data grounding; (b) selects the minimum sufficient repair scope based on error location and dependency scope; and (c) compiles category-level repair knowledge and instance-specific evidence into executable contracts for reliable local repair, followed by artifact-level auditing and safe commit.

\noindent\textbf{(3) Consistent improvements across diverse optimization-modeling generators:} We evaluate six benchmarks, four generation pipelines, four fine-tuned generators, and three repair backbones. Accuracy improves by 6.3--20.1 percentage points across the four generation pipelines. Three complementary ablations examine the roles of dual-evidence detection, evidence-guided repair granularity, and contract-constrained local repair.

\section{Related Work}
\noindent\textbf{General LLM hallucination detection and mitigation.}
Hallucination research studies factual errors and deviations from source content~\citep{ji2023survey}. FActScore assesses source support for atomic claims~\citep{factscore}, while SelfCheckGPT uses sampling-based consistency for detection~\citep{selfcheckgpt}. RARR revises unsupported content while retaining the original text where possible~\citep{rarr}, and CRITIC uses tool feedback for correction~\citep{critic}. However, claim-level grounding and output consistency alone do not establish complete requirement coverage or fidelity across mathematical models and solver code. \method addresses these issues through cross-artifact checks grounded in problem requirements and instance data.

\noindent\textbf{Generation and training for LLM-based optimization modeling.}
Chain-of-Experts, OptiMUS, ORMind, and OptiTree organize generation through expert collaboration, feedback, counterfactual reasoning, or tree search; NEMO adds optimizer--simulator validation~\citep{coe,optimus,ormind,optitree,nemo}. ORLM, LLMOPT, ReSocratic, and OptMATH improve training through customizable instructions, multi-element modeling, inverse construction, and bidirectional synthesis~\citep{orlm,llmopt,optibench,optmath}. These designs center on generating correct models and code, but do not by themselves establish that the resulting artifacts are faithful to the original problem. Reliable post-processing repair is therefore needed to assess problem fidelity and correct affected artifacts while preserving valid content.

\noindent\textbf{Verification and correction for LLM-based optimization modeling.}
SIRL improves modeling through solver-informed learning, while ORLoopBench studies solver-in-the-loop diagnosis and repair~\citep{sirl,orloopbench}. StepORLM combines solver outcomes with process supervision~\citep{steporlm}; SAC-Opt uses semantic anchors for code correction~\citep{sacopt}; and Opt-Verifier combines structural and solution verification~\citep{optverifier}. Existing work primarily treats modeling errors as failures to be avoided during generation or corrected within refinement workflows, while scope-aware repair that preserves correct content remains underexplored. In contrast, we study repair as a post-hoc transformation of frozen artifacts, where unnecessary edits to initially correct model--code pairs are themselves failures. \method selects repair granularity based on affected model and code locations, specifies local edit boundaries, coupled updates, and preservation conditions, and audits candidates before acceptance.

\section{Methodology}
\label{sec:method}
\subsection{Problem Formulation and Available Information}
Given a natural-language optimization problem $P$, an upstream generator produces a mathematical model $M_0$ and solver code $S_0$. We check whether the model and code are faithful to $P$. Unsupported additions, omitted requirements, incorrect formulations, model--code inconsistencies, data-binding faults, and result-extraction errors count as hallucinations. Equivalent forms remain valid.

\method takes problem $P$ and the upstream generator's artifacts $M_0$ and $S_0$ as input, and returns $(M^*,S^*)$. It aims to rescue incorrect artifacts while preserving correct content. Its goal can be summarized as follows: after processing by \method, the model is faithful to $P$, the code faithfully implements the model, and both use the instance data correctly. Domain requirements come only from $P$. Detection reads fixed $P/M_0/S_0$ and evidence from $P$. It excludes upstream solve status, error messages, objective values, and reference answers. Execution checks follow candidate generation; references are used only for offline evaluation.

\begin{figure}[t]
  \centering
  \includegraphics[width=\linewidth]{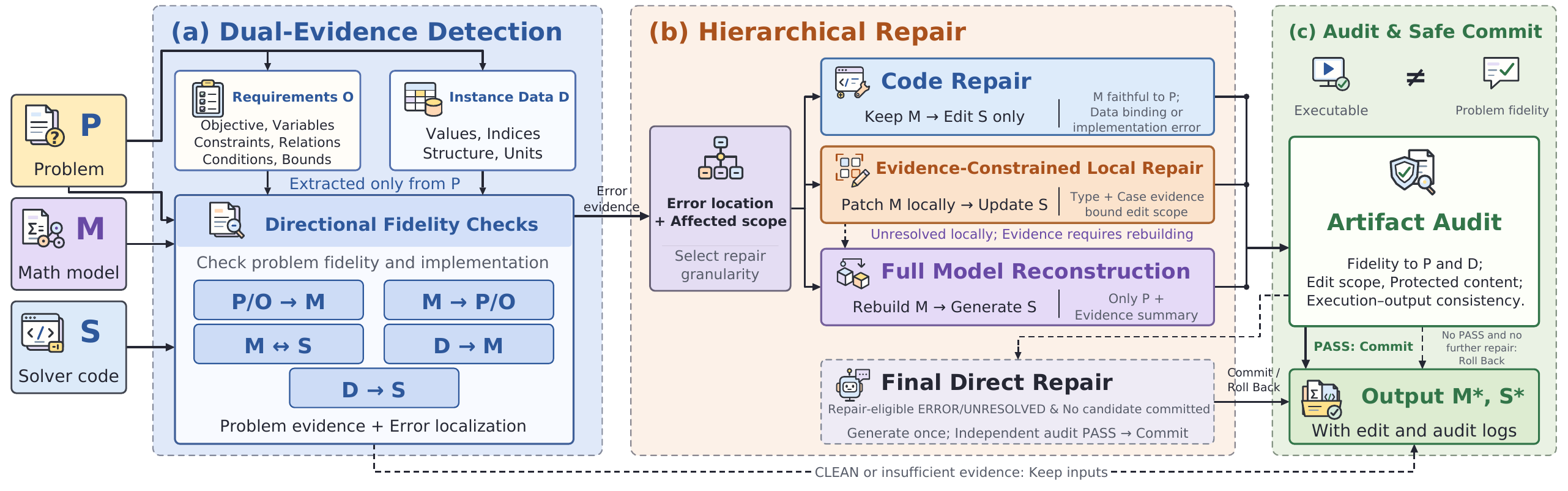}
  \caption{Overview of the \method workflow.}
  \label{fig:overview}
\end{figure}

Figure~\ref{fig:overview} shows the \method workflow: directional checks produce grounded findings; affected scope and evidence determine repair granularity; contracts bound local edits. All candidates undergo artifact auditing before commit; \clean inputs are retained.

\subsection{Dual Evidence and Directional Fidelity Checks}
Requirements $O$ specify decision semantics; instance data $D$ bind values, indices, keys, and units. These distinct evidence spaces support complementary directional fidelity checks, evaluated jointly in one TraceDetect LLM request and then validated deterministically (Appendix~\ref{app:operational-realization}).

\paragraph{Requirements $O$: what the problem specifies.}
Extract source-grounded requirements:
\begin{equation}
  O=\{(\mathrm{id}_i,\mathrm{role}_i,\mathrm{requirement}_i,
       \mathrm{source}_i,\mathrm{deps}_i)\}_{i=1}^{m}.
\end{equation}
Each requirement cites source text for objectives, domains, relations, conditions, quantifiers, or bounds. $O_{\mathrm{rel}}$ links them via shared variables, indices, or conditions without adding requirements.

\paragraph{Instance data $D$: the data to which requirements apply.}
$D$ records values and structure, parsed deterministically when explicit or extracted from $P$ and checked against it. Separating $O$ and $D$ exposes binding errors, such as correct profits assigned to the wrong products.

\paragraph{Cross-checking the problem, model, and code.}
These checks constitute a directional decomposition of artifact fidelity, not repeated critiques of the same output. Each direction defines a distinct verification obligation and exposes a complementary failure surface (Table~\ref{tab:directional-checks}).

\begin{table}[!htb]
  \centering
  \caption{Directional decomposition of fidelity checks.}
  \label{tab:directional-checks}
  \begin{minipage}{\linewidth}
    \centering
    \small
    \setlength{\tabcolsep}{3pt}
    \hyphenpenalty=10000
    \renewcommand{\arraystretch}{1.15}
    \begin{tabular*}{\linewidth}{@{}p{0.116\linewidth}@{\hspace{3pt}}>{\raggedright\arraybackslash}p{0.4074\linewidth}@{\hspace{3pt}}>{\raggedright\arraybackslash}p{\dimexpr0.4766\linewidth-6pt\relax}@{}}
    \toprule
    Direction & Verification obligation & Failure modes exposed \\
    \midrule
    $(P,O)\!\to\!M$ & Complete and faithful requirement coverage & Omitted requirements, reversed relations, incorrect quantifiers \\
    $M\!\to\!(P,O)$ & Source support for every model element & Unsupported variables, constraints, or assumptions \\
    $M\!\leftrightarrow\!S$ & Faithful model-to-code implementation & Model--code construction discrepancies \\
    $D\!\to\!M$ & Correct grounding of data in model objects & Incorrect values, indices, units, or bindings \\
    $D\!\to\!S$ & Correct materialization of data in code & Incorrect keys, sets, arrays, or index instantiation \\
    \bottomrule
    \end{tabular*}
  \end{minipage}
\end{table}

Grounded outputs are merged by root cause into localized findings with evidence, observed and intended expressions, target references, and dependency scopes:
\begin{equation}
  f=(e_P,\mathrm{observed},\mathrm{intended},\mathrm{refs},\mathrm{deps}).
  \label{eq:finding}
\end{equation}
\textsc{MergeRoot} groups findings $\mathcal F_j$ from direction $j$ by component, normalized target, and divergence kind, deduplicating support:
\begin{equation}
  \begin{aligned}
    \mathcal G &= \operatorname{MergeRoot}\!\left(\textstyle\bigcup_{j\in\mathcal J}\mathcal F_j\right),\\
    \mathcal J &= \{(P,O)\!\to\!M,\ M\!\to\!(P,O),\ M\!\leftrightarrow\!S,\ D\!\to\!M,\ D\!\to\!S\}.
  \end{aligned}
  \label{eq:root-merge}
\end{equation}
Different targets remain separate; conflicting effects withhold local writes. Detection returns \clean (no error found), \err (supported error), or \unres (uncertain). For initial errors, non-\clean counts toward Error Coverage, whereas grounded findings guide granularity; non-\clean alone never authorizes edits. Appendix~\ref{app:zero-shot-screening} summarizes evidence-sufficiency routing.

\subsection{Evidence-Guided Repair Granularity Selection}
\label{sec:hierarchy}
For root cause $g$, $\Gamma_g$ contains targets requiring coordinated changes, including source-grounded missing elements; $\operatorname{Refs}(S)$ contains code locations. A local write set is \emph{closed} when supported, nonconflicting edits to existing targets cover all required dependent updates without rebuilding $M$. With $\mathrm{S\text{-}only}\prec\mathrm{M\text{-}local}\prec\mathrm{Structural}$, select the smallest route whose closed write set covers $\Gamma_g$:
\begin{equation}
  r(g)=
  \begin{cases}
    \mathrm{S\text{-}only}, & \Gamma_g\subseteq\operatorname{Refs}(S),\ \text{no identified error in $M$},\\
    \mathrm{M\text{-}local}, & \text{a model error with a closed local write set covering $\Gamma_g$},\\
    \mathrm{Structural}, & \text{evidence rules out a closed local write set covering $\Gamma_g$},\\
    \mathrm{Unresolved}, & \text{evidence does not support the preceding decisions}.
  \end{cases}
  \label{eq:repair-routing}
\end{equation}
Thus, granularity is not a free LLM choice: TraceDetect proposes a scope and route, but deterministic evidence, reference, and edit-permission checks gate authorization under this minimum-sufficient-scope policy. Minimum is relative to supported routes and evidenced dependencies. Structural requires evidence of local insufficiency: missing variables or variable families, new index structures, missing constraint families, or coupled changes across model components. Uncertainty alone yields Unresolved, not permission to enlarge the scope.

S-only preserves $M$; M-local patches $M$ and affected $S$. Structural rebuilds from $P/O/D$ without old $M/S$. Unresolved alone grants no repair; absent further evidence, artifacts are retained. Local failure never authorizes reconstruction; Section~\ref{sec:commit} specifies the audited final-direct fallback.

\subsection{Evidence-Constrained Local Repair}
\label{sec:local}
\paragraph{Repair cards and instance tasks.}
A \emph{RepairCard} specifies applicability, required evidence, allowed and prohibited edits, and acceptance criteria for one of 24 error categories. A \emph{RepairTask} binds these rules to source text, observed and intended expressions, targets, and dependencies (Equation~\eqref{eq:finding}). Figure~\ref{fig:local} shows this combination. A concrete example and its corresponding code fields are provided in Appendix~\ref{app:ratio-illustration}.

\begin{figure}[H]
  \centering
  \includegraphics[width=\linewidth]{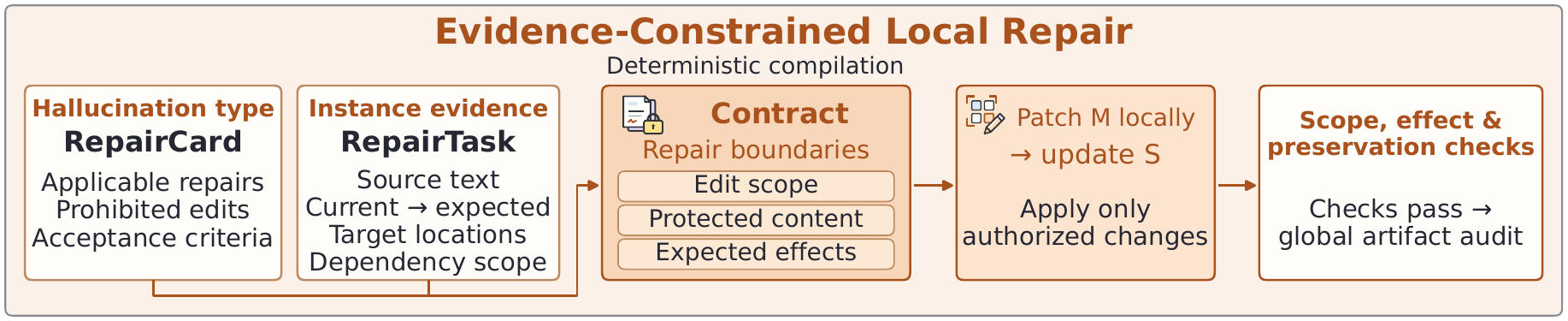}
  \caption{Evidence-constrained local repair for the M-local branch.}
  \label{fig:local}
\end{figure}

\paragraph{Deterministic contract compilation.}
A fixed compiler converts the RepairCard and RepairTask into executable edit restrictions and acceptance rules. These form a \emph{Contract}:
\begin{equation}
  \mathcal{C}=(W,R,E,K,V),
  \label{eq:contract}
\end{equation}
Here, $W$ contains writable locations and $R$ contains read-only dependencies used for checking. $E$ specifies intended effects, $K$ the content to preserve, and $V$ the acceptance rules. For a locally closed task, $T_g$ contains the error locations. $\mathcal E_g$ records dependencies supported by $P$ that require coordinated edits. Starting from $T_g$, we add locations that must change together:
\begin{equation}
  W^{(0)}=T_g,\qquad
  W^{(k+1)}=W^{(k)}\cup
  \{v\mid u\in W^{(k)},\ (u,v)\in\mathcal E_g\},\qquad
  W=W^{(k_*)}.
  \label{eq:write-closure}
\end{equation}
We stop when no new locations are added: $W^{(k_*+1)}=W^{(k_*)}$. A dependency used only for checking belongs to $R$; referencing it does not grant permission to edit it. All updates must preserve $K$. Changes to domains, objective meaning, or unrelated constraints need their own evidence from $P$.

\paragraph{Applying and checking a patch.}
The repairer proposes $\Delta M$ to produce $M'$. Deterministic checks enforce the edit scope $W$ and preserve $K$. Evidence checks verify intended effects:
\begin{equation}
  \operatorname{supp}(\Delta M)\subseteq W,
  \qquad M'|_K=M|_K,
  \qquad V(M',E)=\mathrm{PASS}.
  \label{eq:patch}
\end{equation}
After the patch passes, we synchronize and check $S$. Diffs and stable identifiers enforce scope and preservation. The example changes only $\texttt{ratio.sense}:\ {\leq}\to{\geq}$, yielding $5B\geq2A$ while preserving coefficients, profits, capacity, and domains.

\subsection{Artifact-Level Auditing and Safe Commit}
\label{sec:commit}
To \emph{commit} is to replace retained artifacts with an approved candidate. Execution checks whether code runs and returns the required format. Artifact-level auditing compares the candidate with $P/O/D$ for problem fidelity, model--code agreement, data binding, result extraction, edit scope, and preservation. It thus evaluates the complete artifact rather than objective value or solver status alone. Variable-domain changes require support from $P$ and comparison with the fixed input. After a failed audit, the framework tries further authorized tasks or keeps the current artifacts.

Let $z_t=(M_t,S_t)$ be the retained artifacts and $\hat z_t=(\hat M_t,\hat S_t)$ a candidate. We commit only when $\operatorname{Exec}$ confirms execution and output format and $\operatorname{Audit}$ confirms evidence and preservation:
\begin{equation}
  z_{t+1}=\hat z_t\ \text{if }\operatorname{Exec}(\hat S_t)=\mathrm{PASS}\land\operatorname{Audit}(P,O,D,z_t,\hat z_t)=\mathrm{PASS};\quad z_{t+1}=z_t\ \text{otherwise}.
  \label{eq:safe-commit}
\end{equation}
Generating a candidate does not replace the retained artifacts. Approved corrections contribute Rescue; preservation checks and artifact-level auditing control Regression. A detection alarm alone does not permit an edit. Absent a controlled commit, one final direct repair is allowed only for remaining \err or evidence-supported \unres instances. We generate this candidate from $P$ and the retained $M/S$, then audit it independently. Every path follows Equation~\eqref{eq:safe-commit}: execution, problem fidelity, and edit constraints must all pass before commit.

\section{Experiments}
\label{sec:experiments}
\subsection{Setup and Evaluation Metrics}
\label{sec:settings}
\paragraph{Benchmarks and paired evaluation.}
We evaluate on MAMO Complex, IndustryOR, ComplexOR, OptMATH, NLP4LP, and BWOR, covering compound constraints, industrial applications, and long-form data. Before-and-after comparisons use fixed $P/M_0/S_0$ from each upstream method. We include usable $M/S$ and follow the established reference-conflict exclusions.

Zero-shot denotes a pipeline that generates a model and then code, without worked examples. The generation frameworks are ORMind, OptiMUS-0.3, and SAC-Opt. The fine-tuned generators are ORLM-LLaMA-3-8B, SIRL-Qwen2.5-7B-Gurobi, LLMOPT-Qwen2.5-14B, and OptMATH-Qwen2.5-7B. DeepSeek V4 Flash repairs both groups; it, Qwen3.8-27B, and GPT-5.6 Sol also repair self-generated zero-shot outputs. Gurobi is the solver.

\paragraph{Metrics.}
Accuracy checks objective values or required solve status, with absolute and relative tolerances of $10^{-2}$ and $10^{-4}$. Let $\mathcal{E}_0$ and $\mathcal{C}_0$ denote outputs scored incorrect and correct, respectively, before repair. Rescue is the fraction of $\mathcal{E}_0$ made correct, Regression is the fraction of $\mathcal{C}_0$ made incorrect, and Error Coverage is the fraction of $\mathcal{E}_0$ assigned a non-\clean screening outcome and thus retained for further examination or potential repair. Repair eligibility for a non-\clean output depends on evidence and routing. Objective/status labels score outcomes, not the broader hallucination space of requirement, formulation, model--code, data-binding, and result-extraction errors; thus, outcome correctness need not imply complete modeling-artifact fidelity (Appendix~\ref{app:result-fidelity} quantifies this disagreement). Overall results aggregate paired instances per upstream. Cross-upstream means equally weight the eight methods in Tables~\ref{tab:trainfree} and~\ref{tab:finetuned}; all values are in \%.

\subsection{Post-Processing Generation Frameworks and Fine-Tuned Models}
\label{sec:upstreams}
\begin{table}[!htb]
  \centering
  \caption{DeepSeek V4 Flash repair of train-free pipelines (\%). Top: benchmark and overall accuracy. MAMO, Industry, and Complex denote MAMO Complex, IndustryOR, and ComplexOR.}
  \label{tab:trainfree}
\fontsize{8.3pt}{9.3pt}\selectfont
\setlength{\tabcolsep}{4pt}
\begin{tabular}{@{}llrrrrrrr@{}}
\toprule
Upstream & Setting & MAMO & Industry & Complex & OptMATH & NLP4LP & BWOR & Overall \\
\midrule
\multirow{2}{*}{Zero-shot} & Initial & 25.1 & 50.6 & 11.8 & 23.5 & 73.6 & 54.9 & 44.8 \\
 & +\method & \textbf{55.5} & \textbf{72.3} & \textbf{70.6} & \textbf{48.8} & \textbf{80.6} & \textbf{67.1} & \textbf{64.9} \\
\midrule
\multirow{2}{*}{ORMind} & Initial & 49.3 & 60.2 & 52.9 & 46.4 & 74.0 & 61.0 & 58.6 \\
 & +\method & \textbf{56.9} & \textbf{66.3} & \textbf{64.7} & \textbf{50.0} & \textbf{79.3} & \textbf{72.0} & \textbf{64.9} \\
\midrule
\multirow{2}{*}{OptiMUS-0.3} & Initial & 39.8 & 48.2 & 47.1 & 33.7 & 77.7 & 46.3 & 51.7 \\
 & +\method & \textbf{58.3} & \textbf{53.0} & \textbf{64.7} & \textbf{39.2} & \textbf{81.4} & \textbf{53.7} & \textbf{60.4} \\
\midrule
\multirow{2}{*}{SAC-Opt} & Initial & 36.0 & 47.0 & 52.9 & 21.7 & 79.3 & 40.2 & 48.1 \\
 & +\method & \textbf{54.0} & \textbf{57.8} & \textbf{64.7} & \textbf{28.3} & \textbf{81.8} & \textbf{48.8} & \textbf{57.2} \\
\bottomrule
\end{tabular}
\par\smallskip
\begin{tabular*}{0.90\linewidth}{@{\extracolsep{\fill}}lrrrr@{}}
\toprule
Overall metric & Zero-shot & ORMind & OptiMUS-0.3 & SAC-Opt \\
\midrule
Rescue $\uparrow$ & \textbf{36.9} & \textbf{16.6} & \textbf{19.1} & \textbf{18.5} \\
Regression $\downarrow$ & 0.6 & 0.9 & 1.0 & 1.0 \\
Error Coverage $\uparrow$ & 94.1 & 97.8 & 98.0 & 98.2 \\
\bottomrule
\end{tabular*}

\end{table}
\begin{table}[!htb]
  \centering
  \caption{DeepSeek V4 Flash repair of fine-tuned generators (\%). Layout follows Table~\ref{tab:trainfree}; model names abbreviate the generators in Section~\ref{sec:settings}.}
  \label{tab:finetuned}
\fontsize{8.3pt}{9.3pt}\selectfont
\setlength{\tabcolsep}{4pt}
\begin{tabular}{@{}llrrrrrrr@{}}
\toprule
Upstream & Setting & MAMO & Industry & Complex & OptMATH & NLP4LP & BWOR & Overall \\
\midrule
\multirow{2}{*}{ORLM-8B} & Initial & 35.5 & 38.6 & 52.9 & 9.0 & 67.4 & 20.7 & 38.8 \\
 & +\method & \textbf{46.9} & \textbf{44.6} & \textbf{52.9} & \textbf{20.5} & \textbf{71.1} & \textbf{35.4} & \textbf{47.4} \\
\midrule
\multirow{2}{*}{SIRL-7B} & Initial & 67.8 & 42.1 & 52.9 & 34.0 & 74.3 & 27.3 & 56.6 \\
 & +\method & \textbf{69.8} & \textbf{59.2} & \textbf{58.8} & \textbf{37.9} & \textbf{75.1} & \textbf{34.8} & \textbf{60.7} \\
\midrule
\multirow{2}{*}{LLMOPT-14B} & Initial & 50.7 & 30.1 & 47.1 & 13.4 & 67.8 & 23.2 & 43.2 \\
 & +\method & \textbf{55.5} & \textbf{42.2} & \textbf{58.8} & \textbf{20.7} & \textbf{69.8} & \textbf{34.1} & \textbf{50.9} \\
\midrule
\multirow{2}{*}{OptMATH-7B} & Initial & 26.1 & 20.5 & 17.6 & 19.3 & 67.4 & 9.8 & 34.7 \\
 & +\method & \textbf{32.2} & \textbf{30.1} & \textbf{35.3} & \textbf{24.1} & \textbf{71.5} & \textbf{30.5} & \textbf{43.4} \\
\bottomrule
\end{tabular}
\par\smallskip
\begin{tabular*}{0.90\linewidth}{@{\extracolsep{\fill}}lrrrr@{}}
\toprule
Overall metric & ORLM-8B & SIRL-7B & LLMOPT-14B & OptMATH-7B \\
\midrule
Rescue $\uparrow$ & \textbf{14.7} & \textbf{10.4} & \textbf{14.3} & \textbf{13.8} \\
Regression $\downarrow$ & 1.0 & 0.7 & 0.9 & 0.8 \\
Error Coverage $\uparrow$ & 100.0 & 100.0 & 100.0 & 96.6 \\
\bottomrule
\end{tabular*}

\end{table}

\method improves all four pipelines (Table~\ref{tab:trainfree}). Zero-shot accuracy rises from 44.8\% to 64.9\%, with 36.9\% Rescue; ORMind, OptiMUS-0.3, and SAC-Opt gain 6.3, 8.7, and 9.1 percentage points. Regression stays at or below 1.0\% throughout. Zero-shot improves consistently across all six benchmarks, with gains ranging from 7.0 percentage points on NLP4LP to 58.8 points on ComplexOR.

Complementing the Zero-shot outcome gains, blind paired manual review of 339 selected DeepSeek V4 Flash Zero-shot artifacts shows source-level artifact fidelity rising from 2.7\% to 51.8\% among 330 adjudicable pairs, with 50.8\% Artifact Rescue. Appendix~\ref{app:result-fidelity} details the cohort, adjudication protocol, and paired transitions.

For fine-tuned generators (Table~\ref{tab:finetuned}), ORLM, SIRL, LLMOPT, and OptMATH gain 8.6, 4.1, 7.7, and 8.7 percentage points; Rescue ranges from 10.4\% to 14.7\%. Post-processing repair complements domain training across model families and sizes.

\subsection{Comparison with Generic Repair Baselines}
\label{sec:prompt}
Prompt-based Repair (Prompt) is instructed to check the same $P/M_0/S_0$ for hallucinations, repair errors, and preserve correct content, returning complete $M/S$ in one response. It receives no repair examples, $O/D$, execution feedback, or references. We compare its gains with \method's multi-stage detection, repair, and audit.

\begin{table}[!htb]
  \centering
  \caption{Repair comparisons (\%). (a) Cross-backbone comparison on each backbone's frozen zero-shot outputs. (b) Controlled comparison of repair baselines on 801 frozen DeepSeek V4 Flash zero-shot outputs. Bold marks \method's final accuracy and Rescue.}
  \label{tab:backbones}
  \begin{minipage}[t]{0.575\linewidth}
    \centering
    \textbf{(a) Cross-backbone comparison}\par\smallskip
\fontsize{7.2pt}{8.2pt}\selectfont
\setlength{\tabcolsep}{2pt}
\begin{tabular}{@{}llrrrrr@{}}
\toprule
Backbone & Repairer & Initial & Final & Rescue & Regr. & \shortstack{Error\\Coverage} \\
\midrule
DeepSeek V4 Flash & Prompt & 44.8 & 56.4 & 21.3 & 0.3 & 50.5 \\
 & \method & 44.8 & \textbf{64.9} & \textbf{36.9} & 0.6 & 94.1 \\
\midrule
Qwen3.8-27B & Prompt & 26.8 & 29.6 & 4.6 & 2.3 & 69.8 \\
 & \method & 26.8 & \textbf{38.2} & \textbf{15.9} & 0.9 & 93.6 \\
\midrule
GPT-5.6 Sol & Prompt & 51.2 & 51.7 & 7.9 & 6.6 & 86.4 \\
 & \method & 51.2 & \textbf{65.7} & \textbf{29.9} & 0.0 & 95.3 \\
\bottomrule
\end{tabular}

  \end{minipage}\hfill
  \begin{minipage}[t]{0.405\linewidth}
    \centering
    \textbf{(b) Controlled comparison of repair baselines}\par\smallskip
\fontsize{6.8pt}{7.8pt}\selectfont
\setlength{\tabcolsep}{1.4pt}
\begin{tabular}{@{}lrrrrr@{}}
\toprule
Repairer & Initial & Final & Rescue & Regr. & \shortstack{Error\\Coverage} \\
\midrule
\multicolumn{6}{@{}l}{\textit{DeepSeek V4 Flash}} \\[-1pt]
Prompt-based & 44.8 & 56.4 & 21.3 & 0.3 & 50.5 \\
CoT-based & 44.8 & 57.7 & 25.6 & 2.8 & 73.0 \\
Multi-stage Prompt & 44.8 & 55.8 & 26.7 & 8.4 & 79.0 \\
\method & 44.8 & \textbf{64.9} & \textbf{36.9} & 0.6 & 94.1 \\
\bottomrule
\end{tabular}

  \end{minipage}
  \par\smallskip
  \begin{minipage}{\linewidth}\footnotesize
  Prompt emits no detection state; changes to $M/S$ in a valid output record that an error was acted on.
  \end{minipage}
\end{table}

\paragraph{Cross-backbone comparison.}
\method improves accuracy and Rescue across all three backbones with low Regression, showing consistent gains across backbones.

\paragraph{Controlled workflow comparison.}
Among the two stronger generic baselines, only Multi-stage matches \method in logical calls and newly processed tokens. Single-call CoT checks requirements, the model, and model--code consistency before returning a complete repair. Multi-stage separates diagnosis, repair, and independent audit, revising after a failed audit. Appendix~\ref{app:enhanced-repair-prompts} provides key prompts and complete call flow; exact templates are in the supplementary code.

Both enhanced baselines improve Rescue and Error Coverage but also introduce more Regression. In contrast, \method achieves 64.9\% final accuracy, 36.9\% Rescue, and only 0.6\% Regression while covering 94.1\% of initially incorrect outputs with a non-\clean screening outcome. Thus, its advantage cannot be explained solely by longer reasoning or more stages: dual-evidence diagnosis, evidence-guided granularity selection, and contract-constrained repair are key to balancing error recovery with correct-content preservation.

\subsection{Repair-Path Contributions Across Upstream Error Structures}
\begin{figure}[!htbp]
  \centering
  \includegraphics[width=\linewidth]{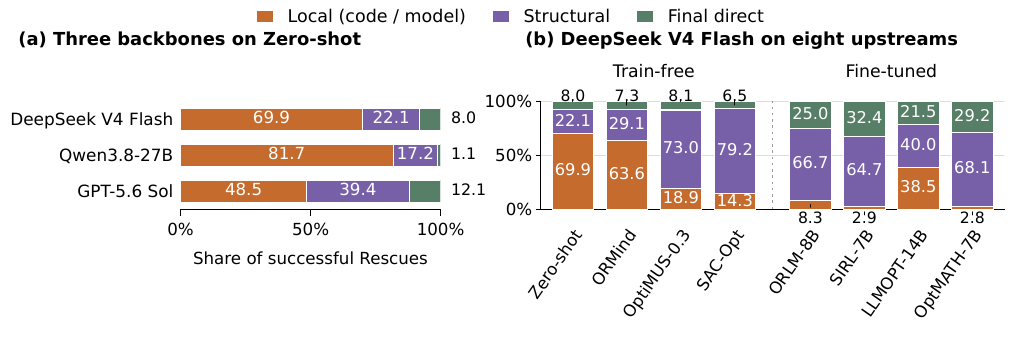}
  \caption{Successful Rescue paths across backbones (a) and upstream generators (b). Local includes S-only and M-local; final direct is the audited fallback. Shares are conditional on successful Rescue.}
  \label{fig:contributions}
\end{figure}
Figure~\ref{fig:contributions} shows that repair granularity varies across input sources. Local repair accounts for 69.9\%, 81.7\%, and 48.5\% of Rescues for DeepSeek, Qwen, and GPT, and over 60\% for Zero-shot and ORMind. Structural repair reaches 73.0\%/79.2\% for OptiMUS-0.3/SAC-Opt and 64.7\%--68.1\% for ORLM, SIRL, and OptMATH. Final direct repair adds 21.5\%--32.4\% for fine-tuned generators.

This variation is consistent with different affected scopes and coupling: a fixed preference for local repair does not represent the successful paths of all upstreams. It supports treating evidence-guided granularity selection as a mechanism for accommodating heterogeneous errors. These are successful-path shares, not routing accuracy or a local-only counterfactual; ablation B below provides complementary evidence for the hierarchy together with its branch evidence and permissions.

\FloatBarrier
\subsection{Mechanism-Wise Ablation Study}
\label{sec:ablation}
The three ablations map one-to-one to the problems and mechanisms in Sections~3.2--3.4: evidence-grounded diagnosis, sufficient-granularity selection, and repair-side-effect control; all use the same frozen DeepSeek zero-shot outputs. \textbf{Detection mechanism (Full vs A):} A replaces $O/D$ and the directional fidelity checks with direct detection. \textbf{Granularity-selection mechanism (Full vs B):} B keeps Full's detection trigger but replaces route-specific repair with one generic $M/S$ repair, removing granularity routing, branch evidence, and edit permissions as an operational package. \textbf{Contracted-repair mechanism (Full vs C):} C replaces only contract-based local repair with generic local rewriting; all other components and budgets remain fixed.

\begin{table}[!htb]
  \centering
  \caption{Mechanism-wise ablation study with DeepSeek V4 Flash (\%). All configurations share the initial outputs. Bold highlights the complete framework.}
  \label{tab:ablation}
\small
\setlength{\tabcolsep}{4pt}
\begin{tabular}{@{}lrrrrr@{}}
\toprule
Configuration & Initial & Final & Rescue $\uparrow$ & Regr. $\downarrow$ & Error Coverage $\uparrow$ \\
\midrule
Full \method & 44.8 & \textbf{64.9} & \textbf{36.9} & 0.6 & 94.1 \\
\midrule
A: w/o dual-evidence diagnosis & 44.8 & 50.2 & 12.4 & 3.3 & 87.1 \\
B: w/o granularity selection & 44.8 & 55.7 & 20.6 & 1.1 & 94.1 \\
C: w/o contracted local repair & 44.8 & 57.8 & 24.4 & 1.1 & 94.1 \\
\bottomrule
\end{tabular}

\end{table}

\paragraph{Mechanism-wise evidence.}
Full vs A yields the largest loss: final accuracy, Rescue, and Error Coverage fall by 14.7, 24.5, and 7.0 points, while Regression rises by 2.7 points. Under B, final accuracy and Rescue fall by 9.2 and 16.3 points; under C, by 7.1 and 12.5 points, supporting granularity selection and contracted local repair. The screening outcome for Error Coverage is finalized before hierarchical repair; hence B and C leave it unchanged at 94.1\%, each raising Regression by 0.5 points. Appendix~\ref{app:sacopt-local} gives signed changes. Other paths partly compensate for the local-repair gap in the system; the fixed-entry ablation isolates it, yielding a 20.0-point Rescue gain.

\begin{figure}[!htb]
  \centering
  \includegraphics[width=0.96\linewidth]{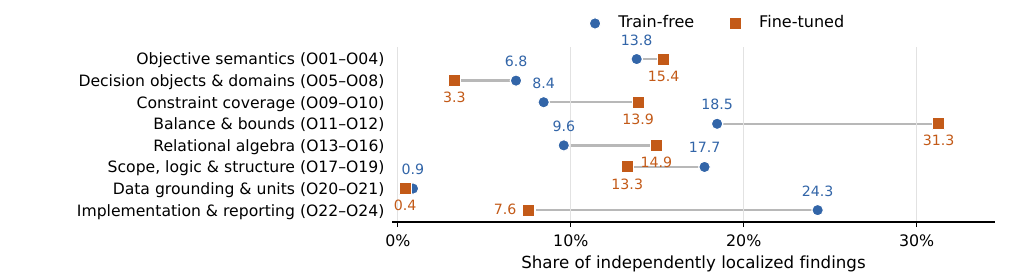}
  \caption{Hallucination distribution across eight taxonomy groups.}
  \label{fig:hallucination-distribution}
\end{figure}
Figure~\ref{fig:hallucination-distribution} profiles 688 train-free and 950 fine-tuned pre-repair findings assigned to the frozen taxonomy, normalized within each family; one problem may yield multiple findings. Implementation and reporting account for 24.3\% of train-free findings. For fine-tuned outputs, balance and bounds account for 31.3\%, and relational algebra for 14.9\%. The profile spans formulation, data grounding, code implementation, and result reporting.
\FloatBarrier

\section{Conclusion}
\method combines dual-evidence diagnosis, evidence-guided granularity selection, and contract-constrained repair with artifact-level auditing and safe commit. By treating hallucination repair as an independent post-hoc transformation of frozen problem--model--code artifacts, \method makes error correction and content preservation joint requirements of optimization-modeling repair. Across six benchmarks, \method improves accuracy and Rescue while maintaining broad Error Coverage and low Regression, moving LLM-based optimization modeling beyond a sole focus on initial success toward evidence-grounded hallucination screening and risk-controlled artifact repair.

\label{page:main-end}
\clearpage
\section*{Reproducibility Statement}
\label{sec:reproducibility-statement}
Sections~\ref{sec:method} and~\ref{sec:experiments} specify the frozen problem--model--code interface, source-grounded directional fidelity checks, evidence-controlled repair hierarchy, contracts, evaluation protocol, and metrics. Appendix~\ref{app:contracts} provides the 24-category RepairCard catalog and concrete RepairTask, compiled Contract, and code-form repair artifacts; Appendix~\ref{app:prompts} documents the baseline prompts and the operational realization of detection, routing, contracts, audit, and commit; Appendix~\ref{app:detailed-results} reports outcome-level and paired source-fidelity results, resource usage, and ablations; Appendix~\ref{app:hallucination-classification} reports the O01--O24 taxonomy and error profiles; and Appendix~\ref{app:repair-output} presents an end-to-end repair record, candidate patch, final model and solver code, execution, and audit output. The code repository at \url{https://anonymous.4open.science/r/OPT-Kraken} releases the solver-independent \method post-processor, including the operational definitions of all 24 RepairCards, source-grounded diagnostic and repair prompts, evidence validation, route and contract compilation, artifact-audit and commit logic, input/output schemas, an OpenAI-compatible command-line interface, a runnable example, and tests.

\label{page:references-start}
\bibliography{references}
\bibliographystyle{plainnat}
\label{page:refs-end}
\clearpage
\appendix
\raggedbottom

\definecolor{artifactgray}{RGB}{245,245,245}
\definecolor{artifactline}{RGB}{48,48,48}
\lstdefinestyle{artifact}{
  basicstyle=\ttfamily\fontsize{7.0pt}{8.0pt}\selectfont,
  backgroundcolor=\color{artifactgray},
  frame=single,
  frameround=ffff,
  rulecolor=\color{artifactline},
  framesep=5pt,
  xleftmargin=0pt,
  xrightmargin=0pt,
  breaklines=true,
  breakatwhitespace=false,
  columns=fullflexible,
  keepspaces=true,
  showstringspaces=false,
  aboveskip=0pt,
  belowskip=8pt
}
\lstdefinestyle{artifactcompact}{
  style=artifact,
  basicstyle=\ttfamily\fontsize{6.2pt}{7.0pt}\selectfont,
  framesep=4pt,
  belowskip=5pt
}
\newcommand{\artifactheading}[1]{%
  \par\noindent\fcolorbox{black}{black}{%
    \parbox{\dimexpr\linewidth-2\fboxsep-2\fboxrule\relax}{%
      \color{white}\bfseries\footnotesize #1}}%
  \par\nobreak\vspace{-\fboxrule}}
\newcommand{\apmetric}[2]{\shortstack{#1\\[-1pt]{\fontsize{5pt}{5.3pt}\selectfont(#2\%)}}}
\makeatletter
\newenvironment{classificationtable}{%
  \par\vspace{5pt}\noindent\begin{minipage}{\linewidth}\def\@captype{table}%
}{\end{minipage}\par\vspace{5pt}}
\makeatother

\begin{center}
  {\Large\bfseries Appendix for Fix What's Wrong, Keep What's Right:\\
  Hallucination Repair for LLM-Based Optimization Modeling}
\end{center}
\vspace{1.2em}
{\large\bfseries TABLE OF CONTENTS}\par
\vspace{0.8em}
\begingroup
\small
\setlength{\parskip}{3pt}
\noindent\textbf{A\quad RepairCards, RepairTasks, and Compiled Contracts}\dotfill\pageref{app:contracts}\par
\hspace*{2em}A.1\quad Operational RepairCard catalog\dotfill\pageref{app:card-catalog}\par
\hspace*{2em}A.2\quad A formal-run objective-repair artifact\dotfill\pageref{app:formal-artifact}\par
\hspace*{2em}A.3\quad RepairTask and compiled Contract\dotfill\pageref{app:task-contract}\par
\hspace*{2em}A.4\quad Code-form ratio-repair trace\dotfill\pageref{app:ratio-illustration}\par
\hspace*{2em}A.5\quad RepairCard field quick reference\dotfill\pageref{app:repair-card-quick-reference}\par
\noindent\textbf{B\quad Prompt Protocols and Operational Realization}\dotfill\pageref{app:prompts}\par
\hspace*{2em}B.1\quad Zero-shot generation\dotfill\pageref{app:zero-shot-prompt}\par
\hspace*{2em}B.2\quad Prompt-based repair\dotfill\pageref{app:repair-prompt}\par
\hspace*{2em}B.3\quad CoT and multi-stage prompt-based repair\dotfill\pageref{app:enhanced-repair-prompts}\par
\hspace*{2em}B.4\quad Detection, routing, contracts, and audit\dotfill\pageref{app:operational-realization}\par
\noindent\textbf{C\quad Detailed Experimental Results}\dotfill\pageref{app:detailed-results}\par
\hspace*{2em}C.1\quad Outcome-Level Correctness and Modeling-Artifact Fidelity\dotfill\pageref{app:result-fidelity}\par
\hspace*{2em}C.2\quad Outcome-level repair and preservation statistics\dotfill\pageref{app:zero-shot-screening}\par
\hspace*{2em}C.3\quad Benchmark-wise Rescue rates\dotfill\pageref{app:benchmark-rescue-radars}\par
\hspace*{2em}C.4\quad Training-free generation pipelines\dotfill\pageref{app:trainfree-detail}\par
\hspace*{2em}C.5\quad Qwen3.8-27B zero-shot generation\dotfill\pageref{app:qwen-zero-shot-detail}\par
\hspace*{2em}C.6\quad Model calls, token usage, and runtime\dotfill\pageref{app:resource-usage}\par
\hspace*{2em}C.7\quad Mechanism-wise and fixed-entry ablations\dotfill\pageref{app:sacopt-local}\par
\noindent\textbf{D\quad Hallucination Taxonomy and Error Profiles}\dotfill\pageref{app:hallucination-classification}\par
\hspace*{2em}D.1\quad DeepSeek V4 Flash across eight upstreams\dotfill\pageref{app:hallucination-deepseek}\par
\hspace*{2em}D.2\quad Qwen and GPT Zero-shot outputs\dotfill\pageref{app:hallucination-backbones}\par
\hspace*{2em}D.3\quad Outcome-conditioned O01--O24 composition\dotfill\pageref{app:hallucination-outcomes}\par
\noindent\textbf{E\quad End-to-End Repair Output}\dotfill\pageref{app:repair-output}\par
\hspace*{2em}E.1\quad Hallucination record and candidate patch\dotfill\pageref{app:detected-output}\par
\hspace*{2em}E.2\quad Final mathematical model and solver code\dotfill\pageref{app:final-output}\par
\hspace*{2em}E.3\quad Execution, artifact audit, and commit\dotfill\pageref{app:audit-output}\par
\endgroup

\section{RepairCards, RepairTasks, and Compiled Contracts}
\label{app:contracts}

\subsection{Operational RepairCard Catalog}
\label{app:card-catalog}
A RepairCard is a reusable, category-level specification of when a repair is applicable, what evidence it requires, which operators it authorizes, what it protects, and how its effect is checked. The catalog covers mathematical formulation, data binding, model-to-code translation, and result extraction. Table~\ref{tab:appendix-repaircards} compactly indexes all 24 operational card categories used by \method. A full inline expansion would require the eligibility schemas, operation templates, protected fields, and executable predicates for every card, together with the TraceDetect and artifact-audit prompts and validator source. We therefore expand O02 as a readable end-to-end instance below; the anonymous code repository cited in the Reproducibility Statement releases the complete operational catalog and the solver-independent post-processing implementation.

\begingroup
\fontsize{7.4pt}{8.5pt}\selectfont
\setlength{\tabcolsep}{3pt}
\setlength{\LTpre}{4pt}
\setlength{\LTpost}{4pt}
\begin{longtable}{@{}p{0.055\linewidth}p{0.105\linewidth}p{0.49\linewidth}p{0.25\linewidth}@{}}
\caption{Compact catalog of the 24 operational RepairCard categories and their admissible contract families.}\label{tab:appendix-repaircards}\\
\toprule
ID & Artifact & Hallucination / repair type & Contract family \\
\midrule
\endfirsthead
\multicolumn{4}{l}{\tablename~\thetable\ (continued)}\\
\toprule ID & Artifact & Hallucination / repair type & Contract family \\
\midrule\endhead
\midrule\multicolumn{4}{r}{Continued on next page}\\\endfoot
\bottomrule\endlastfoot
O01 & P$\rightarrow$M / M & Objective Goal or Sense & objective\_global \\
O02 & P$\rightarrow$M / M & Objective Term, Coefficient, or Support & objective\_local \\
O03 & P$\rightarrow$M / M & Objective Structure, Risk, or Multiobjective Rule & objective\_structure \\
O04 & P$\rightarrow$M / M & Hard/Soft Requirement Role & hard\_soft\_role \\
O05 & P$\rightarrow$M / M & Decision Object, Family, or Index & variable\_structure \\
O06 & P$\rightarrow$M / M & Continuous-to-Discrete Domain & domain\_to\_discrete \\
O07 & P$\rightarrow$M / M & Discrete-to-Continuous Domain & domain\_to\_continuous \\
O08 & P$\rightarrow$M / M & Variable Bound, Sign, or Allowed Set & variable\_bound \\
O09 & P$\rightarrow$M / M & Missing Required Constraint or Family & missing\_constraint \\
O10 & P$\rightarrow$M / M & Unsupported Constraint or Family & unsupported\_constraint \\
O11 & P$\rightarrow$M / M & Equality, Balance, or Conservation Form & equality\_balance \\
O12 & P$\rightarrow$M / M & Scalar Bound, Threshold, or Strictness & scalar\_boundary \\
O13 & P$\rightarrow$M / M & Pairwise Comparative or Multiple Relation & pairwise\_comparative \\
O14 & P$\rightarrow$M / M & Share of a Combined Total & total\_share \\
O15 & P$\rightarrow$M / M & Ratio to Another Quantity & other\_ratio \\
O16 & P$\rightarrow$M / M & General Row Algebra or Role Binding & row\_algebra \\
O17 & P$\rightarrow$M / M & Index, Quantifier, or Aggregation Scope & index\_scope \\
O18 & P$\rightarrow$M / M & Logic, Activation, or Linearization & logic\_activation \\
O19 & P$\rightarrow$M / M & Network, Temporal, Scheduling, or Global Structure & global\_structure \\
O20 & P$\rightarrow$M / M & Parameter Value, Key, or Presence & parameter\_grounding \\
O21 & P$\rightarrow$M / M & Unit, Scale, Time, or Applicability Scope & unit\_time\_scope \\
O22 & M$\rightarrow$S / S & Mathematical-Model-to-Code Construction & code\_construction \\
O23 & M$\rightarrow$S / S & Code Index or Set Materialization & code\_index\_materialization \\
O24 & S$\rightarrow$R / S & Solve, Extraction, or Reporting & result\_protocol \\
\end{longtable}

\endgroup

\subsection{A Formal-Run Objective-Repair Artifact}
\label{app:formal-artifact}
We use \texttt{industryor:prob\_085} as a compact end-to-end example. The problem asks for monthly profit from two pet-food products. Source-grounded unit contribution is
\[
  c_1=2.80-0.25-2(0.20)-3(0.50)=0.65,
  \qquad
  c_2=2.00-0.20-3(0.20)-1.5(0.50)=0.45.
\]
The generated model instead simplified the same expressions to $1.15x_1+0.65x_2$. Dual-evidence checking therefore emits an \err finding localized to the objective. O02 applies because the profit metric and objective direction are correct, while two source-bound coefficients are inconsistent.

\artifactheading{RepairCard O02 --- Objective Term, Coefficient, or Support}
\begin{lstlisting}[style=artifact]
{
  "type_id": "O02",
  "name": "Objective Term, Coefficient, or Support",
  "applicability_boundary": "Use O01 when the base metric itself changes. Use O20 only when an auditable parameter object is already wrong before it reaches the objective; correct data bound to the wrong term belong here.",
  "repair_goal": "Insert, delete, deduplicate, or rebind only the divergent terms; never replace price-minus-cost by gross price merely because a selling-price value is easier to quote.",
  "required_evidence": [
    {
      "source": "P",
      "description": "Exact authoritative source evidence: Build a source-to-objective ledger containing semantic event, variable, coefficient formula, sign, unit, and index support; when price and variable cost are both supplied, distinguish gross revenue from net contribution and retain fixed charges separately."
    },
    {
      "source": "M",
      "description": "The current localized model target and its dependency context."
    },
    {
      "source": "P_AND_M",
      "description": "Evidence that applies this operational boundary: Use O01 when the base metric itself changes. Use O20 only when an auditable parameter object is already wrong before it reaches the objective; correct data bound to the wrong term belong here."
    }
  ],
  "forbidden_actions": [
    "Do not invent a source fact, value, set, weight, business rule, or output requirement.",
    "Do not use any evaluation-only artifact outside the current P/M/S/R runtime interface.",
    "Do not edit outside the localized target and proved dependency closure.",
    "Do not force a type or patch when a required instance binding is unresolved.",
    "Do not treat solver success, infeasibility, unboundedness, or a surprising value alone as semantic evidence."
  ],
  "allowed_operators": [
    "INSERT_OBJECTIVE_TERM",
    "DELETE_OBJECTIVE_TERM",
    "DEDUPLICATE_OBJECTIVE_TERM",
    "SET_OBJECTIVE_TERM_SIGN",
    "REBIND_OBJECTIVE_COEFFICIENT",
    "REPLACE_OBJECTIVE_INDEX_DOMAIN",
    "ALIGN_OBJECTIVE_TIME_INDEX"
  ],
  "acceptance_templates": [
    {
      "predicate_id": "OBJECTIVE_TERM_COVERAGE",
      "required_bindings": [
        "p_requirement",
        "m_target",
        "semantic_contract"
      ],
      "criterion": "Hand-evaluate the complete repaired objective on a fixed feasible decision, including variable and fixed costs, and perturb one affected variable to verify its exact marginal contribution."
    },
    {
      "predicate_id": "OBJECTIVE_COEFFICIENT_MATCH",
      "required_bindings": [
        "p_requirement",
        "m_target",
        "semantic_contract"
      ],
      "criterion": "Hand-evaluate the complete repaired objective on a fixed feasible decision, including variable and fixed costs, and perturb one affected variable to verify its exact marginal contribution."
    },
    {
      "predicate_id": "EDIT_SCOPE_VALID",
      "required_bindings": [
        "changed_refs",
        "writable_refs"
      ],
      "criterion": "Only the evidence-bound target and proved dependency closure may change."
    },
    {
      "predicate_id": "AFFECTED_TRACE_PASS",
      "required_bindings": [
        "problem_evidence",
        "repaired_artifact"
      ],
      "criterion": "The repaired local trace must agree with unchanged authoritative evidence."
    },
    {
      "predicate_id": "EXECUTION_NONREGRESSION",
      "required_bindings": [
        "pre_execution_signature",
        "post_execution_signature"
      ],
      "criterion": "The repair must not introduce a parse, build, solve, or extraction failure."
    }
  ]
}
\end{lstlisting}

\subsection{RepairTask and Compiled Contract}
\label{app:task-contract}
The RepairTask binds O02 to the source spans, observed coefficients, intended coefficients, and stable target reference. Deterministic compilation converts that instance binding into an atomic replacement with explicit preconditions. The Contract exposes the exact writable set---the objective and its two affected terms---and freezes every variable and constraint.

\artifactheading{RepairTask Materialized from the Formal Run}
\begin{lstlisting}[style=artifactcompact]
{
  "task_id": "T-INDUSTRYOR-085-OBJECTIVE",
  "requirement_ids": [
    "R-83E296E071",
    "R-97EABD0CFB",
    "R-5B68B6A5CB",
    "R-6E9E3006BB",
    "R-9BBB56BB2E"
  ],
  "component": "objective",
  "target_refs": [
    "objective"
  ],
  "divergence": "INCONSISTENT",
  "type_id": "O02",
  "source_evidence": "Meaties/Yummies sell for $2.80/$2.00 per pack; grain costs $0.20/lb, meat costs $0.50/lb, and mixing/packaging costs $0.25/$0.20 per pack.",
  "diagnosis": "M0 simplifies the source-grounded profit arithmetic incorrectly: the coefficients 1.15 and 0.65 must be 0.65 and 0.45.",
  "observed_fact": {
    "sense": "max",
    "constant": -0.0,
    "coefficients": {
      "x1": 1.15,
      "x2": 0.65
    }
  },
  "intended_fact": {
    "sense": "max",
    "constant": 0.0,
    "coefficients": {
      "x1": 0.65,
      "x2": 0.45
    }
  },
  "operation": "replace_objective",
  "allowed_scope": "ATOMIC"
}
\end{lstlisting}
\artifactheading{Compiled Contract}
\begin{lstlisting}[style=artifactcompact,basicstyle=\ttfamily\fontsize{5.9pt}{6.5pt}\selectfont,framesep=3pt]
{
  "task_id": "T-INDUSTRYOR-085-OBJECTIVE",
  "type_id": "O02",
  "allowed_scope": "ATOMIC",
  "operations": [
    {
      "op": "replace_objective",
      "observed": {
        "sense": "max",
        "constant": -0.0,
        "coefficients": {
          "x1": 1.15,
          "x2": 0.65
        }
      },
      "value": {
        "sense": "max",
        "constant": 0.0,
        "coefficients": {
          "x1": 0.65,
          "x2": 0.45
        }
      }
    }
  ],
  "writable_refs": [
    "objective",
    "objective_term:x1",
    "objective_term:x2"
  ],
  "freeze_all_other_model_elements": true
}
\end{lstlisting}

\clearpage
\subsection{Code-Form Ratio-Repair Trace}
\label{app:ratio-illustration}
Figure~\ref{fig:repair-code} expands the running example in Figure~\ref{fig:example} as a code-form trace instantiated with the actual runtime classes; it is an illustrative mechanism example, not an additional benchmark result. The trace exposes the applicable O15 rules and acceptance predicates, the source-bound RepairTask, the compiled atomic operation and writable set, and the post-application checks. The complete constraint row appears before and after compilation: only its relation changes from $\leq$ to $\geq$, while the coefficients, right-hand side, and all other model elements remain protected.
\begin{figure}[htbp]
  \centering
  \begin{minipage}{0.98\linewidth}
\begin{lstlisting}[
    style=artifactcompact,
    language=Python,
    basicstyle=\ttfamily\fontsize{7.2pt}{8.2pt}\selectfont,
    numbers=left,
    numberstyle=\tiny\color{gray},
    numbersep=6pt,
    xleftmargin=16pt,
    framexleftmargin=14pt,
    framesep=4pt
  ]
# Selected fields from the runtime-instantiated O15 example.
repair_card = {
    "type_id": "O15",
    "required_proportion_basis": "other_ratio",
    "allowed_operators": [
        "REPLACE_CONSTRAINT_SEMANTICS",
        "REBIND_CONSTRAINT_SUBSCRIPTS",
    ],
    "acceptance_predicates": [
        "PROPORTION_SEMANTICS_MATCH",
        "EDIT_SCOPE_VALID",
        "AFFECTED_TRACE_PASS",
        "EXECUTION_NONREGRESSION",
    ],
}

repair_task = {
    "type_id": "O15",
    "source_evidence": (
        "B's output must be at least 2/5 of A's output."
    ),
    "target_refs": ["constraint:ratio"],
    "diagnosis": (
        "The inequality reverses the source-grounded at-least ratio."
    ),
    "observed_fact": {
        "name": "ratio", "coefficients": {"B": 5, "A": -2},
        "sense": "<=", "rhs": 0,
    },
    "intended_fact": {
        "name": "ratio", "coefficients": {"B": 5, "A": -2},
        "sense": ">=", "rhs": 0,
    },
    "operation": "upsert_constraint",
    "allowed_scope": "ATOMIC",
    "dual_supported": True,
}

compiled_contract = {
    "operations": [{
        "op": "upsert_constraint",
        "observed": repair_task["observed_fact"],
        "value": repair_task["intended_fact"],
    }],
    "writable_refs": ["constraint:ratio"],
    "freeze_all_other_model_elements": True,
}

application = {
    "status": "applied",
    "checks": {
        "direction_corrected": True,
        "all_other_model_content_preserved": True,
    },
}
\end{lstlisting}
  \end{minipage}
  \caption{Code-form trace from RepairCard and RepairTask to Contract application for the illustrative O15 ratio repair.}
  \label{fig:repair-code}
\end{figure}

\clearpage
\subsection{RepairCard Field Quick Reference}
\label{app:repair-card-quick-reference}
To give readers a rapid overview of the RepairCard catalog, the listing below projects each O01--O24 category onto four descriptive fields: \texttt{type\_id}, \texttt{name}, \texttt{applicability\_boundary}, and \texttt{repair\_goal}. This is an explanatory documentation view rather than a verbatim specification of the executable repair logic. The implementation includes additional evidence schemas, operator restrictions, protected fields, acceptance predicates, uncertainty handling, and code-level branching, and some representations or wording therefore differ from this appendix projection. The released implementation governs runtime behavior; this subsection is provided only as a compact conceptual reference.

\artifactheading{Illustrative O01--O24 RepairCard Field Projection}
\begin{lstlisting}[
  style=artifactcompact,
  basicstyle=\ttfamily\fontsize{5.2pt}{5.9pt}\selectfont,
  framesep=3pt
]
[
  {
    "type_id": "O01",
    "name": "Objective Goal or Sense",
    "applicability_boundary": "Use O02 when the global goal is correct and only a local term, coefficient, sign, or support is wrong. Use O03 for multiobjective, risk, fixed-charge, piecewise, nonlinear, or fractional objective structure. Minimizing $f$ and maximizing $-f$ are equivalent only when the required reported value is mapped back correctly.",
    "repair_goal": "Correct the global sense or metric without rewriting already-correct local terms."
  },
  {
    "type_id": "O02",
    "name": "Objective Term, Coefficient, or Support",
    "applicability_boundary": "Use O01 when the base metric itself changes. Use O20 only when an auditable parameter object is already wrong before it reaches the objective; correct data bound to the wrong term belong here.",
    "repair_goal": "Insert, delete, deduplicate, or rebind only the divergent terms; never replace price-minus-cost by gross price merely because a selling-price value is easier to quote."
  },
  {
    "type_id": "O03",
    "name": "Objective Structure, Risk, or Multiobjective Rule",
    "applicability_boundary": "Use O01 for a plain global sense or base-metric error and O02 for local linear term errors. Do not invent weights, priorities, risk levels, breakpoints, or approximations that $P$ does not specify.",
    "repair_goal": "Restore an exact formulation block or return `UNSURE_MISSING_EVIDENCE`."
  },
  {
    "type_id": "O04",
    "name": "Hard/Soft Requirement Role",
    "applicability_boundary": "A source-authorized elastic constraint, aspiration level, or epsilon constraint is valid. Preference language alone does not authorize inventing a weight, priority, or tolerance.",
    "repair_goal": "Restore the hard rule or remove the false hardening; introduce a soft mechanism only when specified."
  },
  {
    "type_id": "O05",
    "name": "Decision Object, Family, or Index",
    "applicability_boundary": "Use O06/O07 for a type-direction error and O08 for bounds on an otherwise correct variable. Use O09 when the variables are sufficient and a restriction over them is missing.",
    "repair_goal": "Add, redefine, re-index, split, merge, or couple only the affected family."
  },
  {
    "type_id": "O06",
    "name": "Continuous-to-Discrete Domain",
    "applicability_boundary": "Integer-valued data and generic nouns such as item, trip, worker, or machine are insufficient by themselves. An explicit request for an LP is affirmative evidence against discretization unless $P$ also explicitly requires whole, integer, binary, indivisible, or selected/not-selected decisions.",
    "repair_goal": "Change only the proved variables to integer or binary."
  },
  {
    "type_id": "O07",
    "name": "Discrete-to-Continuous Domain",
    "applicability_boundary": "The absence of discreteness words is not proof of continuity. A phrase such as amount, quantity, or allocation is insufficient unless the problem or formulation class affirmatively supports divisibility.",
    "repair_goal": "Relax only those domains to continuous."
  },
  {
    "type_id": "O08",
    "name": "Variable Bound, Sign, or Allowed Set",
    "applicability_boundary": "Use O12 when the mismatch is a modeled scalar constraint row rather than the variable domain, and O20 when the bound parameter was extracted incorrectly before binding. Never infer a bound from solver unboundedness alone.",
    "repair_goal": "Change only the evidenced bound/domain member."
  },
  {
    "type_id": "O09",
    "name": "Missing Required Constraint or Family",
    "applicability_boundary": "Use O10 for an unsupported extra rule; O11 for an existing equality/balance replaced by the wrong form; O12--O17 when an intended row exists but its semantic signature is wrong; and O18/O19 when a coupled formulation block is required. Infeasibility or unboundedness alone never proves a missing rule.",
    "repair_goal": "Insert the complete missing row/family only."
  },
  {
    "type_id": "O10",
    "name": "Unsupported Constraint or Family",
    "applicability_boundary": "A redundant valid inequality is not a hallucination. Before deleting a row, disprove every source role it may serve, including demand, capacity, budget, minimum, maximum, balance, and contract satisfaction.",
    "repair_goal": "Delete only the unsupported row or family after proving no required role is lost."
  },
  {
    "type_id": "O11",
    "name": "Equality, Balance, or Conservation Form",
    "applicability_boundary": "Use O12 for a pure scalar threshold/boundary error, O13 for an ordinary pairwise comparison, and O16 for wrong coefficients or entity roles in an otherwise correctly formed row. Use O09 when the balance family is wholly absent. O11 requires explicit equality/balance provenance or a current equality target whose unsupported side is being removed. \"The rest,\" \"using all available capacity,\" an ordinary demand statement, or a participation verb does not by itself authorize equality.",
    "repair_goal": "Restore the sourced equality/complete two-sided form, or replace an unsupported equality by the sourced one-sided relation."
  },
  {
    "type_id": "O12",
    "name": "Scalar Bound, Threshold, or Strictness",
    "applicability_boundary": "Use O13 for a pairwise comparison or multiple, O14/O15 for percentages or fractions, O16 for coefficient/role changes, and O08 when the value belongs directly to a variable-domain bound. Under the standard LP/MILP convention, strict natural-language comparisons--including infix forms such as \"more X than Y\"--use the corresponding non-strict solver boundary unless the source supplies a positive numeric margin; do not invent epsilon or a one-unit lattice offset.",
    "repair_goal": "Correct only the grounded direction, value, or boundary representation; retain the standard non-strict boundary when no margin is given."
  },
  {
    "type_id": "O13",
    "name": "Pairwise Comparative or Multiple Relation",
    "applicability_boundary": "This type excludes fractions or percentages: use O14 when a group is a share of the combined total and O15 when it is a fraction of another named quantity. Use O12 for comparison with a scalar threshold and O11 only for an equality/balance form. A strict pairwise phrase without a stated positive margin retains the standard non-strict LP/MILP boundary.",
    "repair_goal": "Compile the canonical row $X-kY\\ \\{<=,>=\\}\\ 0$ and replace only the affected row."
  },
  {
    "type_id": "O14",
    "name": "Share of a Combined Total",
    "applicability_boundary": "Use O15 when the source says $G$ is a fraction of another separately named quantity $O$, not of the combined total. Use O13 for non-fractional multiples. The business quantity may be count, tons, capacity, cost, or another sourced unit; variable counts must not silently replace weighted quantities.",
    "repair_goal": "Compile $(1-\\alpha)G-\\alpha O\\ge0$ for at least or $\\le0$ for at most, using source-supported term weights. A source-validated canonical row supersedes the obsolete same-name scalar or reversed row, but never an independent companion bound."
  },
  {
    "type_id": "O15",
    "name": "Ratio to Another Quantity",
    "applicability_boundary": "Use O14 when the denominator is the combined total $G+O$. Use O13 for factors outside fractional/share language or ordinary pairwise order. Do not infer the denominator from variable names.",
    "repair_goal": "Compile $G-\\alpha O\\ge0$ for at least or $\\le0$ for at most."
  },
  {
    "type_id": "O16",
    "name": "General Row Algebra or Role Binding",
    "applicability_boundary": "Use O12--O15 for their dedicated scalar/comparative/proportion signatures and O17 when the main defect is index, quantifier, or aggregation scope. Use O20 only when the parameter object itself is wrong before binding.",
    "repair_goal": "Rebuild only that row or homogeneous row family."
  },
  {
    "type_id": "O17",
    "name": "Index, Quantifier, or Aggregation Scope",
    "applicability_boundary": "Use O05 when the variable family itself lacks an axis. Use O16 when the axes are correct and only coefficients or roles inside one row are wrong. Use O19 when a global structural invariant, rather than an indexed linear family, is missing.",
    "repair_goal": "Replace the affected index domain or complete the row family only when one consistent ledger is available."
  },
  {
    "type_id": "O18",
    "name": "Logic, Activation, or Linearization",
    "applicability_boundary": "Use O09 only when a directly stated independent row is missing. O18 applies when correctness requires a coupled truth-table or companion-row block. A Big-M value is not wrong merely because it is loose unless validity or a claimed guarantee changes.",
    "repair_goal": "Restore the smallest complete logical block; partial repair is forbidden."
  },
  {
    "type_id": "O19",
    "name": "Network, Temporal, Scheduling, or Global Structure",
    "applicability_boundary": "Use O17 for a local index/aggregation mistake in an otherwise correct structure and O18 for a bounded logical block. O19 requires reasoning over multiple rows, states, periods, nodes, or activities.",
    "repair_goal": "Restore the smallest complete structural block supported by $P$ only when every participating variable, row family, and boundary is bound. A broadly collapsed or disconnected parsed model with an unbounded dependency closure is reported as `UNRESOLVED`; it does not authorize an automatic Full-M rewrite."
  },
  {
    "type_id": "O20",
    "name": "Parameter Value, Key, or Presence",
    "applicability_boundary": "Correct data attached to the wrong objective term, constraint role, or variable belong to O02, O08, or O16. Do not attribute an earlier extraction error when no auditable parameter object exists.",
    "repair_goal": "Correct the ledger entry or key mapping only."
  },
  {
    "type_id": "O21",
    "name": "Unit, Scale, Time, or Applicability Scope",
    "applicability_boundary": "Different objective and constraint coefficients may legitimately have different units. Use O20 for a wrong raw value/key and O02/O16 when correctly scaled data are bound to the wrong component.",
    "repair_goal": "Correct only the proved conversion or scope."
  },
  {
    "type_id": "O22",
    "name": "Mathematical-Model-to-Code Construction",
    "applicability_boundary": "Use O01--O21 when $M$ is already wrong and $S$ faithfully implements it. Generic syntax, connection, dependency, license, and environment failures are not this type without a localized semantic construction divergence.",
    "repair_goal": "Repair the smallest code construction that diverges."
  },
  {
    "type_id": "O23",
    "name": "Code Index or Set Materialization",
    "applicability_boundary": "Use O17 when the mathematical index contract itself is wrong. Use O22 for a non-indexed construction divergence.",
    "repair_goal": "Correct only the faulty expansion or filter."
  },
  {
    "type_id": "O24",
    "name": "Solve, Extraction, or Reporting",
    "applicability_boundary": "Use O22/O23 when the wrong model was constructed. Termination at an accepted gap is valid when an incumbent is reported without claiming proven optimality.",
    "repair_goal": "Correct only the readout, inverse transformation, rounding, post-processing, or status protocol."
  }
]
\end{lstlisting}

\clearpage
\section{Prompt Protocols and Operational Realization}
\label{app:prompts}
The first three subsections document the generation and generic-repair baselines; the final subsection specifies how the principal method modules are realized. Braced fields are populated per instance. Zero-shot generation first maps $P$ to a complete mathematical model and then maps the frozen model to executable Gurobi code. Prompt-based Repair receives the same $P/M_0/S_0$ tuple and returns one complete repaired model--code pair.

\artifactheading{Core Call Graphs for Zero-Shot and Direct Prompt Repair}
\begin{lstlisting}[style=artifactcompact,language=Python]
# Core baseline protocol code; transport and JSON validation are omitted.
def zero_shot(P):
    model_reply = call_model(
        system=MODEL_SYSTEM_PROMPT,
        user=MODEL_USER_TEMPLATE.format(question=P),
    )
    M = parse_json(model_reply)["mathematical_model"]
    code_reply = call_model(
        system=CODE_SYSTEM_PROMPT,
        user=CODE_USER_TEMPLATE.format(mathematical_model=M),
    )
    S = parse_json(code_reply)["solver_code"]
    return M, S


def prompt_based_repair(P, M0, S0):
    repair_reply = call_model(
        system=PROMPT_REPAIR_SYSTEM,
        user=PROMPT_REPAIR_USER.format(P=P, M0=M0, S0=S0),
    )
    repaired = parse_json(repair_reply)
    return repaired["mathematical_model"], repaired["solver_code"]
\end{lstlisting}

\subsection{Zero-Shot Generation}
\label{app:zero-shot-prompt}
\artifactheading{Zero-Shot Prompts: Stage 1 ($P\rightarrow M$) and Stage 2 ($M\rightarrow S$)}
\begin{lstlisting}[style=artifactcompact]
[Stage 1: P -> M]
SYSTEM
You are an operations research expert. Given a natural-language optimization problem P, produce the corresponding complete mathematical model M.

USER
Optimization problem P:

{question}

Return exactly one JSON object:
{
  "mathematical_model": "the complete mathematical model M"
}

[Stage 2: M -> S]
SYSTEM
You are an operations research expert and Gurobi programmer. Given a mathematical optimization model M, produce the corresponding executable Gurobi solver code S.

USER
Mathematical model M:

{mathematical_model}

Return exactly one JSON object:
{
  "solver_code": "the complete executable Python source S"
}

The solver code must faithfully implement M, use gurobipy, define solve_model() with no arguments, solve the model, and return:
{"status":"optimal|infeasible|unbounded|error", "objective_value":number_or_null, "reported_value":number_or_null, "mip_gap":number_or_null}.
\end{lstlisting}

\subsection{Prompt-Based Repair}
\label{app:repair-prompt}
\artifactheading{Prompt-Based Repair: System Prompt and Per-Case Message}
\begin{lstlisting}[style=artifactcompact]
SYSTEM
You receive an optimization problem P, its current mathematical model M0, and its current Gurobi solver code S0.

Carefully inspect M0 and S0 against P. Check whether M0 or S0 contains any hallucination or error. If there is an error, correct it. If M0 and S0 are already correct, return them unchanged.

Return exactly one JSON object:
{
  "mathematical_model": "complete final mathematical model",
  "solver_code": "complete final Python source code"
}

The returned mathematical_model and solver_code must be complete, even when no correction is needed. The final Python code must use gurobipy, define solve_model() with no arguments, and return:
{"status":"optimal|infeasible|unbounded|error", "objective_value":number_or_null, "reported_value":number_or_null, "mip_gap":number_or_null}.

USER
P -- original optimization problem:
{P}

M0 -- current mathematical model:
{M0}

S0 -- current Gurobi solver code:
```python
{S0}
```

Return the final M and S in the required JSON object.
\end{lstlisting}

\subsection{CoT and Multi-Stage Prompt-Based Repair}
\label{app:enhanced-repair-prompts}
The following prompt excerpts retain the instructions and output fields that determine baseline behavior while omitting repeated formatting boilerplate. CoT-based Repair orders diagnosis and complete repair within one response. Multi-stage Prompt-based Repair separates diagnosis, candidate generation, independent review, and review-triggered revision.

\artifactheading{Key Prompt Excerpts}
\begin{lstlisting}[style=artifactcompact]
[CoT-based Repair: key prompt]
You receive only the original problem P, the current mathematical model M0,
and the current solver code S0. In one response, complete these steps in order:
1. extract the requirements from P;
2. check M0 against those requirements;
3. check whether S0 implements M0;
4. list only confirmed errors; and
5. return a complete corrected model and solver code.
If no error is confirmed, preserve the current semantics.

Do not use or invent an obligation/data ledger, taxonomy, repair route,
contract, write set, solver output, answer, reference model, or external data.
Return exactly one JSON object with fields diagnosis, mathematical_model,
and solver_code. The code must define solve_model(), use gurobipy, and return
status, objective_value, reported_value, and mip_gap.

[Multi-stage Prompt-based Repair: key prompts]
DIAGNOSIS: Compare P with M0 and S0. Do not repair. Return has_error,
a concise diagnosis, and concrete issues with artifact, location, observed,
and expected fields. Use no solver outcome, answer, reference, or framework
artifact.

REPAIR: Given P/M0/S0 and the diagnosis, return one complete candidate model
and solver code. Correct confirmed errors and preserve faithful content.
Choose the edit scope yourself; no route, taxonomy, dependency closure,
contract, protected set, or write permission is provided.

REVIEW: Independently compare the candidate with P and check model-code
consistency. Do not modify or execute it. Return PASS or FAIL, a concise
review, and confirmed issues. Use no answer, reference, or solver output.

REVISION AFTER FAIL: Given P/M0/S0, the diagnosis, current candidate, and
review, return a complete revised model and code. Fix confirmed review issues
while preserving faithful content. Do not return a partial patch.

[Per-case fields]
P:  {original optimization problem}
M0: {current mathematical model}
S0: {current solver code with line numbers}
\end{lstlisting}
\artifactheading{Compact Call Flow}
\begin{lstlisting}[style=artifactcompact,language=Python]
# Compact pseudocode; API, logging, and exception handling are omitted.
def cot_based_repair(P, M0, S0):
    reply = call_model(COT_PROMPT, P, M0, S0)
    diagnosis, candidate = parse_diagnosis_and_complete_pair(reply)
    return execute(candidate)


def multi_stage_prompt_repair(P, M0, S0):
    diagnosis = call_model(DIAGNOSIS_PROMPT, P, M0, S0)
    candidate = call_model(REPAIR_PROMPT, P, M0, S0, diagnosis)
    review = call_model(REVIEW_PROMPT, P, M0, S0, diagnosis, candidate)

    for _ in range(3):
        if review.decision == "PASS":
            break
        candidate = call_model(
            REVISION_PROMPT, P, M0, S0, diagnosis, candidate, review
        )
        review = call_model(
            REVIEW_PROMPT, P, M0, S0, diagnosis, candidate
        )

    return execute(candidate)
\end{lstlisting}

\paragraph{Experimental rules.}
Both enhanced baselines operate on the same 801 frozen DeepSeek V4 Flash Zero-shot $P/M_0/S_0$ tuples, with temperature zero and thinking disabled. They receive no answer, reference model, solver outcome, execution feedback, structured $O/D$, taxonomy, repair route, or contract. CoT makes exactly one model call. Multi-stage always performs diagnosis, repair, and review; each failed review permits one complete revision followed by re-review, for at most three such rounds and nine logical calls, with early stopping after \textsc{Pass}. Every repair or revision must return a complete $M/S$ pair, and the last complete candidate is executed once. Objective/status scoring is applied offline only after generation. For Error Coverage, an initially incorrect CoT output is counted as covered when its ordered diagnosis explicitly identifies an artifact discrepancy; Multi-stage uses the diagnosis-stage \texttt{has\_error} decision. Accordingly, Table~\ref{tab:backbones}(b) operationalizes Error Coverage using each method's native screening signal---diagnosis, \texttt{has\_error}, a non-\clean state, or an acted-on valid repair---and should not be read as a calibrated cross-method detection comparison.

\subsection{Operational Realization of Detection, Routing, and Commit}
\label{app:operational-realization}

\paragraph{Freezing $O$ and $D$.}
The requirement extractor receives $P$ and a deterministic segmentation of its source clauses, but not $M_0$ or $S_0$. Its prompt requests a compact set of JSON requirement records and an assignment for every source clause. Each requirement includes an exact source span, semantic role, entities, optional normalized fact, numeric derivations, and dependency identifiers. A deterministic postprocessor relocates every quoted span in $P$, replaces temporary identifiers with stable ones, validates the component and fact schema, permits only source-grounded percentage, reciprocal, ratio, or complement derivations, and marks uncovered material clauses for review. It then freezes $O$. The relation view $O_{\mathrm{rel}}$ is computed deterministically from dependencies and explicit relation cues already present in the source-grounded requirements; it cannot add a requirement absent from $P$.

For $D$, exact parsers first read literal mappings and Markdown, \LaTeX, or HTML tables. When none applies, at most one auxiliary LLM call receives only $P$ and proposes explicit values, entities, lists, tables, directed pairs, units, axes, and key domains. A fact is retained only if its quote and every literal, unit, axis, and key can be located in that quote. Rejected fields make the manifest incomplete rather than allowing $M_0$ or $S_0$ to fill missing data. Figure~\ref{fig:evidence-extraction} shows the operative prompt fields and the abbreviated call path.

\begin{figure}[!t]
  \artifactheading{Abbreviated $O/D$ Extraction Prompts and Call Path}
\begin{lstlisting}[style=artifactcompact,language=Python]
O_PROMPT = r"""
Extract compact modeling requirements from problem P only. Do not inspect,
diagnose, or repair M or S. Return one JSON object:
{
  "requirements": [{
    "requirement_id": "temporary id",
    "component": "variable|variable_bound|objective|constraint|parameter|output",
    "source_span": "exact contiguous excerpt from P",
    "semantic_role": "short role",
    "entities": ["source entity or intended variable"],
    "normalized_fact": "complete executable fact or null",
    "derivations": ["percentage|reciprocal|ratio|complement record"],
    "dependency_ids": ["temporary id"]
  }],
  "clause_assignments": [{
    "clause_id": "C001",
    "classification": "REQUIREMENT|NON_MODEL_CONTEXT",
    "requirement_ids": ["temporary id"]
  }]
}
Assign every source clause exactly once. Copy every source_span exactly.
Emit normalized_fact only when P fully determines it; otherwise use null.
Never invent a number, domain, variable, constraint, or objective term.
"""

D_PROMPT = r"""
Extract the instance-data manifest D from problem P only. Return one JSON object:
{
  "status": "COMPLETE|INCOMPLETE|NONE",
  "facts": [{
    "name": "stable fact name",
    "kind": "scalar|set|list|table|relation|shape",
    "value": "JSON value copied from P",
    "unit": "explicit unit or null",
    "axes": ["explicit axis"],
    "key_domain": ["explicit key or pair"],
    "source_quote": "exact contiguous excerpt from P"
  }]
}
Preserve values, source order, axes, keys, units, zeros, and pair direction.
If required data are absent or only exemplified, return INCOMPLETE; do not infer
them from M, S, a benchmark template, or common sense.
"""


def freeze_evidence(P, llm_json):
    clauses = segment_source_clauses(P)  # deterministic offsets and clause IDs
    O_raw = llm_json(O_PROMPT, {"P": P, "clauses": clauses})
    O = validate_ground_and_freeze_requirements(P, clauses, O_raw)
    O_rel = deterministic_relation_closure(P, O)

    D = parse_literal_mappings_and_tables(P)
    if D is None:                       # at most one P-only fallback proposal
        D_raw = llm_json(D_PROMPT, {"P": P})
        D = validate_ground_and_freeze_manifest(P, D_raw)
    return O, O_rel, D
\end{lstlisting}
  \caption{The extraction interface. The prompts use $P$ only; all accepted records are subsequently grounded and frozen by deterministic validators.}
  \label{fig:evidence-extraction}
\end{figure}

\paragraph{Directional fidelity checks, root merging, and routing.}
The directional fidelity checks are evaluated jointly within one TraceDetect request, rather than through independent LLM calls. This request jointly compares $(P,O)\!\rightarrow\!M$, $M\!\rightarrow\!(P,O)$, $M\!\leftrightarrow\!S$, $D\!\rightarrow\!M$, and $D\!\rightarrow\!S$. It proposes the strongest source-grounded root, its target references, and its minimum complete write set. The implementation is hybrid: the LLM performs the cross-artifact comparison, while deterministic code validates the response schema, exact source quotes, existing ModelIR references, code-line ranges, algebraic equivalence where it is decidable, and data literals and shapes. Some syntactically decidable discrepancies are emitted directly by rules and enter the same queue without another LLM call.

After validation, \textsc{MergeRoot} uses a precise identity rather than textual similarity. Two findings merge only when their component, normalized stable target, and divergence kind (missing, unsupported, or inconsistent) are identical. Supporting check directions are deduplicated, while findings with different targets or divergence kinds remain separate even when their diagnoses sound similar. If findings under one merge key disagree on the observed or intended effect, the RepairTask compiler records a directional-fact conflict and withholds local write authority. The framework then expands the dependencies of each valid merged root to compute the model and code locations that must change together.

The same TraceDetect call proposes a repair scope and route, but its proposal does not authorize a write. Minimum sufficient scope is the routing policy: semantic scope and dependency identification are LLM-assisted, while deterministic checks enforce source grounding, valid references, and permitted edits. This is not a certificate of global edit minimality or complete semantic dependency discovery. Under this policy, S-only applies only when the model has no supported error and every required write is confined to code or data binding. Local model repair requires existing target references and a complete, nonconflicting local write set. Structural reconstruction requires positive source-grounded evidence that the supported local operators cannot cover the necessary changes, such as missing variables or variable families, a new index structure, missing constraint families, or coupled changes across model components. Failure to establish a local closure alone is not such evidence: absent an independently grounded structural requirement, ambiguous evidence, an unknown target, conflicting writes, or unresolved dependencies produce \unres. Internally, unresolved cases are separated by evidence sufficiency: ULE permits at most one evidence-completion step, whereas ULC retains the current artifacts and does not enter generative repair. Structural, local, and unresolved decisions are therefore LLM-assisted proposals with rule-based final authority.

\paragraph{Contract compilation and artifact audit.}
For a local route, the compiler consumes a RepairCard and a source-bound RepairTask; it does not generate new mathematical meaning. It materializes a finite operation list---for example, replacing an objective, changing a coefficient or variable bound, or inserting, replacing, or deleting an authorized constraint. Each operation carries its observed-value precondition and intended value. The compiled Contract also records writable references, read-only dependencies, intended effects, and a freeze rule for every other model element.

Acceptance predicates with typed operands execute automatically. The current precondition must match; the realized diff must be contained in the writable set; the compiled operation must have its exact effect; frozen references and requirement definitions must remain unchanged; model schemas and references must close; and grounded data shapes, keys, units, and cited code lines must be valid. The implementation also checks that $S$ parses and builds the Gurobi model. Generated-code artifacts are invoked through the required no-argument \texttt{solve\_model()} interface. Committed ModelIR artifacts are executed by \texttt{solve\_quantified\_model} through a public \texttt{solve\_model()} wrapper; both paths return the same status, objective-value, reported-value, and MIP-gap record. A valid infeasible or unbounded status is an execution result, not evidence that the artifact is unfaithful to the problem.

Predicates requiring interpretation of the problem statement are handled by a separate LLM audit. It compares the candidate with $P/O/D$ along objective meaning, variable semantics and domains, constraints and quantifiers, instance-data use, result extraction, and $M$--$S$ agreement, returning \textsc{Pass}, \textsc{Fail}, or \textsc{Unsure}. Deterministic code subsequently validates the audit schema, exact source quotes, candidate references, and agreement between dimension-level and top-level decisions. Variable-domain changes require positive source evidence. On the final direct path, the independent audit also rederives one indirectly specified value or relation from $P$ rather than accepting a candidate literal. Thus, a RepairCard's natural-language acceptance criterion becomes automatic only when the compiler binds it to a typed fact and reference; the remaining problem-fidelity criteria stay explicit audit obligations. Solver objective values and reference answers are never used by this online audit.

\paragraph{Failure handling and bounded retries.}
An audit failure never overwrites the retained pair. The candidate is discarded, after which the next independent, nonconflicting RepairTask may be attempted if it already has complete evidence; the failed root is not resampled. A failed local audit enters structural reconstruction only when write-set analysis independently proves that local scope is insufficient, rather than merely because the first candidate failed.

Each root permits at most one local candidate. Each instance permits at most one fresh structural model, one evidence-grounded refinement of that model, and, after the model is frozen, one code-only repair. If no controlled candidate commits, the framework permits one final direct-repair candidate only for a remaining \err instance or an \unres instance that has obtained sufficient evidence for repair, followed by one independent artifact audit. A \textsc{Fail}, \textsc{Unsure}, invalid interface, or failed deterministic predicate on that path restores the last retained $M/S$ and terminates without a commit. Transport recovery may resume an interrupted, hash-verified stage, but it does not create another repair candidate or relax these budgets.

\FloatBarrier
\section{Detailed Experimental Results}
\label{app:detailed-results}
Initial and Final report Accuracy before and after repair. Let $\mathcal{E}_0$ and $\mathcal{C}_0$ denote outputs scored incorrect and correct, respectively, before repair. Rescue is the fraction of $\mathcal{E}_0$ made correct, Error Coverage is the fraction of $\mathcal{E}_0$ assigned a non-\clean screening outcome, and Regression is the fraction of $\mathcal{C}_0$ made incorrect. These outcome labels follow objective/status scoring; the detector additionally examines formulation, model--code consistency, data-binding, and result-extraction errors. Metric entries in the detailed result tables are percentages. For DeepSeek V4 Flash Zero-shot, among the 442 initially incorrect outputs, 245 (55.4\%) were assigned \err, 171 (38.7\%) were assigned \unres, and 26 (5.9\%) were assigned \clean. The resulting non-\clean Error Coverage is 94.1\%.

\paragraph{Metric interpretation.}
Regression is not a detector false-positive rate: a correct input may receive a non-\clean outcome without Regression if no candidate passes audit and the original is retained. Error Coverage measures coverage of potential repair opportunities, whereas low Regression indicates outcome-level preservation of initially correct outputs. Neither metric measures detection precision or specificity or establishes calibrated binary hallucination detection.

\subsection{Outcome-Level Correctness and Modeling-Artifact Fidelity}
\label{app:result-fidelity}
Following the established benchmark protocols, we evaluate all methods using objective values or the solve status required by each problem to ensure deterministic, reproducible, and consistent evaluation across methods. This criterion is not strictly equivalent to complete fidelity of the modeling artifacts to the original problem: an incorrect modeling element may, for example, be inactive at the optimum of the current instance and leave the reported objective unchanged.

To quantify this discrepancy, we manually reviewed the initial artifacts for all 359 numerically correct DeepSeek V4 Flash Zero-shot results. Each artifact was inspected for substantive inconsistencies between the original problem and the generated model--code artifacts. We found 4 artifacts containing optimization-modeling hallucinations that outcome-level evaluation did not expose: the generated artifacts were not fully faithful to the problem, yet the numerical result was coincidentally correct. The corresponding proportion is 1.1\% (4/359). Table~\ref{tab:appendix-result-fidelity} summarizes the concrete errors and why their numerical effects were masked in these instances.

These observations show that outcome-level correctness and modeling-artifact fidelity are related but non-equivalent. To avoid defining benchmark ground truth through our own detector or manual review, Accuracy, Rescue, and Regression in the main experiments continue to use the same external outcome evaluator for all methods. The manual analysis is used only as a complementary validity check.

\begin{classificationtable}
  \centering
  \caption{Manually reviewed cases of coincidentally correct objective values in DeepSeek V4 Flash Zero-shot outputs.}
  \label{tab:appendix-result-fidelity}
\begingroup
\small
\setlength{\tabcolsep}{3pt}
\renewcommand{\arraystretch}{1.08}
\begin{tabular*}{\linewidth}{@{\extracolsep{\fill}}>{\raggedright\arraybackslash}p{0.18\linewidth}>{\raggedright\arraybackslash}p{0.37\linewidth}>{\raggedright\arraybackslash}p{0.42\linewidth}@{}}
\toprule
Cases & Concrete error & Why the objective remains correct \\
\midrule
\texttt{nlp4lp:}\newline \texttt{52} & Omits the lower bound of $5$ large boxes. & An optimum uses $20$ small and $6$ large boxes, already satisfying the bound. The minimum remains $26$ boxes. \\
\addlinespace[3pt]
\texttt{industryor:}\newline \texttt{prob\_018} & Misreads the conditional minimum of $10$ chairs from supplier B as $10$ orders ($100$ chairs). & The optimum buys only from supplier C, not A, so the condition is not triggered. The minimum cost remains $4000$. \\
\addlinespace[3pt]
\texttt{industryor:}\newline \texttt{prob\_099} & Incorrectly symmetrizes an asymmetric bandwidth matrix, changing reverse-link data. & The changed links do not alter the optimal path $A\to C\to D\to E$ or its bottleneck bandwidth, which remains $84$. \\
\addlinespace[3pt]
\texttt{bwor:}\newline \texttt{BWOR-015} & Reverses shortage and surplus signs in the inventory balance, allowing unmet demand to escape the shortage penalty. & The correct instance also admits a plan meeting demand and capacity requirements with all prioritized penalties equal to $0$; the objective is unchanged. \\
\bottomrule
\end{tabular*}
\endgroup

\end{classificationtable}

\paragraph{Paired pre- and post-repair source fidelity.}
We further examine whether the outcome-level improvements are accompanied by improvements in the fidelity of the generated artifacts to the source problem. We consider the 339 initially outcome-incorrect DeepSeek V4 Flash Zero-shot artifacts selected for evidence-controlled follow-up under the screening policy.

For each instance, both the initial artifact $(P,M_0,S_0)$ and the final retained artifact $(P,M^*,S^*)$ are manually reviewed. Pre- and post-repair artifacts are anonymized and randomly shuffled before review, so annotators are unaware of repair status. The review is conducted by researchers with expertise in operations research, and boundary cases are resolved through adjudication.

Each artifact is assigned one of three source-fidelity labels: \textbf{faithful (F)}, \textbf{unfaithful (U)}, or \textbf{indeterminate (I)}. An artifact is labeled faithful when no substantive inconsistency or unsupported modeling assumption is found relative to the source problem. It is labeled unfaithful when such a discrepancy is present. A case is labeled indeterminate when the source problem is internally inconsistent or does not uniquely determine a material modeling convention, so fidelity cannot be resolved without introducing an external assumption.

Source-fidelity labels are assigned without using the benchmark reference objective or solve-status label. Consequently, source fidelity and outcome correctness need not coincide: a source-faithful artifact may still receive an outcome-incorrect label because of benchmark/reference disagreements, evaluator or interface mismatches, or execution/numerical failures, while outcome correctness may mask modeling errors as illustrated in Table~\ref{tab:appendix-result-fidelity}.

Cases whose fidelity cannot be resolved directly from the source text are adjudicated under the same source-fidelity criterion. Table~\ref{tab:appendix-fidelity-adjudication} summarizes the 15 convention-sensitive or internally ambiguous cases identified during this review.

\begin{classificationtable}
  \centering
  \caption{Adjudication of source-ambiguous or convention-sensitive cases in the paired artifact-fidelity review. F, U, and I denote faithful, unfaithful, and indeterminate, respectively.}
  \label{tab:appendix-fidelity-adjudication}
  \footnotesize
  \setlength{\tabcolsep}{3.2pt}
  \begin{tabular}{@{}p{0.29\linewidth}p{0.12\linewidth}p{0.53\linewidth}@{}}
    \toprule
    Cases & Transition & Adjudication basis \\
    \midrule
    \texttt{optmath:23}, \texttt{optmath:151} & $\mathrm{F}\!\rightarrow\!\mathrm{F}$ (2) & The former faithfully represents an infeasible supply--demand system. In the latter, the source describes quadratic constraints and explicitly declares all variables free; it does not impose the nonnegative axis required by an SOC interpretation. \\
    \texttt{nlp4lp:26} & $\mathrm{U}\!\rightarrow\!\mathrm{U}$ & Replaces the explicit total of 500 with ``at least 500''; infeasibility does not justify relaxing the stated requirement. \\
    \texttt{optmath:47/74/79} & $\mathrm{U}\!\rightarrow\!\mathrm{U}$ (3) & Both initial and final artifacts introduce the unsupported assumption that all units are initially off, materially affecting startup costs. \\
    \texttt{optmath:137/142/}\newline\texttt{148/156} & $\mathrm{I}\!\rightarrow\!\mathrm{I}$ (4) & The source simultaneously invokes SOC semantics and a free-axis variable, producing an unresolved domain inconsistency. \\
    \texttt{nlp4lp:234/82} & $\mathrm{I}\!\rightarrow\!\mathrm{I}$ (2) & The LP instruction and the counting semantics of personnel/pills do not uniquely determine integrality from the source alone. \\
    \texttt{bwor:BWOR-047} & $\mathrm{I}\!\rightarrow\!\mathrm{I}$ & The source does not uniquely specify whether the spare ship first held at the end of the horizon incurs a full year of maintenance and storage cost. \\
    \texttt{industryor:prob\_039} & $\mathrm{I}\!\rightarrow\!\mathrm{I}$ & The source does not uniquely specify whether the four named commodities themselves may be imported. \\
    \texttt{optmath:11} & $\mathrm{I}\!\rightarrow\!\mathrm{I}$ & The target times are introduced as hard ``must land no earlier than'' bounds while the objective also permits and penalizes early landing, leaving their hard/soft role unresolved. \\
    \bottomrule
  \end{tabular}
\end{classificationtable}

The adjudication yields 2 faithful, 4 unfaithful, and 9 indeterminate pairs among these 15 boundary cases. The nine indeterminate pairs remain $\mathrm{I}\!\rightarrow\!\mathrm{I}$ and are excluded from the primary binary fidelity statistics rather than being forced into either fidelity state.

\paragraph{Fidelity metrics.}
Primary fidelity statistics are computed over the \textbf{330 adjudicable pairs}, excluding the nine $\mathrm{I}\!\rightarrow\!\mathrm{I}$ cases. Pre- and post-repair Fidelity denote the fractions of adjudicable artifacts labeled faithful before and after processing. \textbf{Artifact Rescue} measures the fraction of initially unfaithful adjudicable artifacts that become faithful:
\begin{equation}
\mathrm{Artifact\ Rescue} = \frac{\#\{F_0=\mathrm{U},F_*=\mathrm{F}\}}{\#\{F_0=\mathrm{U}\}}.
\label{eq:artifact-rescue}
\end{equation}
Because only nine adjudicable artifacts are initially faithful, preservation is reported through paired transition counts rather than emphasized as a percentage metric.

\begin{classificationtable}
  \centering
  \caption{Paired source-level artifact fidelity analysis on 339 initially outcome-incorrect DeepSeek V4 Flash Zero-shot artifacts selected for follow-up. Primary fidelity statistics are computed over the 330 adjudicable pairs.}
  \label{tab:appendix-artifact-fidelity}
  \begin{tabular}{@{}p{0.68\linewidth}r@{}}
    \toprule
    Metric & Result \\
    \midrule
    Adjudicable pairs & \textbf{330/339} \\
    Pre-repair Fidelity & \textbf{2.7\% (9/330)} \\
    Post-repair Fidelity & \textbf{51.8\% (171/330)} \\
    Artifact Rescue & \textbf{50.8\% (163/321)} \\
    Indeterminate pairs & \textbf{2.7\% (9/339)} \\
    \bottomrule
  \end{tabular}
\end{classificationtable}

Source-level artifact fidelity increases by \textbf{49.1 percentage points}, from 2.7\% (9/330) before repair to 51.8\% (171/330) after processing. Among the 321 initially unfaithful adjudicable artifacts, 163 (50.8\%) are restored to source fidelity.

The corresponding paired transitions are
\begin{equation*}
\begin{aligned}
(\mathrm{U},\mathrm{U})&:158,\qquad (\mathrm{U},\mathrm{F}):163,\\
(\mathrm{F},\mathrm{U})&:1,\qquad (\mathrm{F},\mathrm{F}):8,\qquad
(\mathrm{I},\mathrm{I}):9.
\end{aligned}
\end{equation*}
Thus, 163 initially unfaithful artifacts become faithful, while eight of the nine initially faithful adjudicable artifacts remain faithful. The nine indeterminate pairs remain indeterminate and are excluded from the primary binary fidelity rates.

\paragraph{Sensitivity to indeterminate cases.}
The conclusion is insensitive to how the nine indeterminate pairs are treated. Assigning all nine to the unfaithful state yields 50.4\% post-repair fidelity and 49.4\% Artifact Rescue, whereas assigning all nine to the faithful state yields 53.1\% post-repair fidelity and 50.8\% Artifact Rescue. Thus, the observed improvement is not driven by the treatment of source-ambiguous cases.

This paired analysis complements rather than replaces the outcome-level evaluation. Source-level artifact fidelity evaluates whether the generated model--code artifact is supported by the source problem, whereas the main benchmark metric evaluates the resulting objective value or solve status. The substantial increase in source fidelity therefore provides independent evidence that the outcome-level gains are accompanied by genuine improvements in the generated artifacts themselves.

\subsection{Outcome-Level Repair and Preservation Statistics}
\label{app:zero-shot-screening}
U-Likely-Error (ULE) and U-Likely-Clean (ULC) are evidence-sufficiency subtypes of \unres: ULE requires additional evidence before repair eligibility is determined, whereas ULC conservatively retains the current artifacts.

\begin{table}[!htbp]
  \centering
  \caption{Outcome-level repair performance with supplementary screening and preservation statistics on the 801 frozen DeepSeek V4 Flash Zero-shot artifacts.}
  \label{tab:appendix-deepseek-zero-shot-screening}
  \begin{tabular}{@{}p{0.68\linewidth}r@{}}
    \toprule
    Statistic & Result \\
    \midrule
    Initial Accuracy & 44.8\% \\
    Final Accuracy & \textbf{64.9\%} \\
    Rescue & \textbf{36.9\%} \\
    Regression & \textbf{0.6\%} \\
    Error Coverage & 94.1\% \\
    \midrule
    Initially incorrect: evidence-controlled follow-up (\err{} + ULE) & 76.7\% \\
    \midrule
    Initially correct: \clean & 78.3\% \\
    Initially correct: ULC & 7.0\% \\
    \midrule
    Initially correct: conservative retention (\clean{} + ULC) & 85.2\% \\
    Initially correct: evidence-controlled follow-up (\err{} + ULE) & 14.8\% \\
    \bottomrule
  \end{tabular}

\end{table}

As shown in Table~\ref{tab:appendix-deepseek-zero-shot-screening}, \method improves accuracy from 44.8\% to 64.9\%, rescuing 163 of the 442 initially outcome-incorrect artifacts while regressing only 2 of the 359 initially outcome-correct artifacts. Error Coverage remains high at 94.1\%. Among the 416 initially outcome-incorrect artifacts assigned a non-\clean screening outcome, 339/416 (81.5\%) entered the \err or ULE routes for evidence-controlled follow-up, whereas 77/416 (18.5\%) were conservatively routed to ULC. Thus, the high Error Coverage is not primarily attributable to routing uncertain cases to ULC. Among the initially outcome-correct artifacts, 281 are classified as \clean and 25 are routed to ULC; consequently, 306/359 (85.2\%) fall on the conservative-retention side, while 53/359 (14.8\%) enter \err or ULE for evidence-controlled follow-up. Of these 53 artifacts, 51 remained outcome-correct after the complete pipeline and only 2 regressed, illustrating that evidence-controlled follow-up does not imply a harmful committed edit. These routing statistics provide supplementary evidence of preservation, while Rescue and Regression remain the primary end-to-end measures of repair effectiveness and outcome-level harm.

\subsection{Benchmark-Wise Rescue Rates}
\label{app:benchmark-rescue-radars}
\begin{figure}[H]
  \centering
  \includegraphics[width=0.98\linewidth]{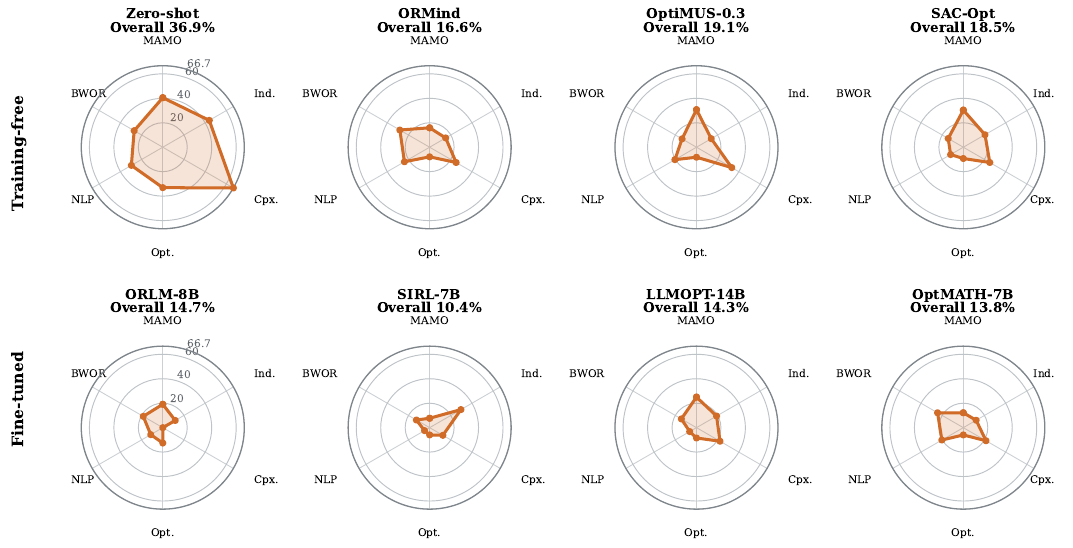}
  \caption{Benchmark-wise Rescue rates for the four training-free pipelines (top) and four fine-tuned generators (bottom). Axis abbreviations denote MAMO Complex, IndustryOR, ComplexOR, OptMATH, NLP4LP, and BWOR. Each panel title gives overall Rescue; the shared outer radius is 66.7\%.}
  \label{fig:benchmark-rescue-radars}
\end{figure}

Figure~\ref{fig:benchmark-rescue-radars} isolates the benchmark-wise Rescue rates underlying the aggregate results in Tables~\ref{tab:trainfree} and~\ref{tab:finetuned}. Each radius is the share of initially outcome-incorrect outputs on that benchmark that become outcome-correct after processing, while each panel title reports the corresponding overall Rescue rate. The top row contains the four training-free pipelines and the bottom row the four fine-tuned generators. All panels share a common scale whose 66.7\% outer ring is the largest observed benchmark-level Rescue rate, enabling direct comparison across upstream methods.

\subsection{Training-Free Generation Pipelines}
\label{app:trainfree-detail}
All results in this subsection use DeepSeek V4 Flash as the repair backbone of \method.
\begingroup
\fontsize{7.7pt}{8.5pt}\selectfont
\setlength{\tabcolsep}{3.4pt}
\setlength{\LTpre}{3pt}
\setlength{\LTpost}{3pt}
\begin{longtable}{@{}lrrrrr@{}}
\caption{Full six-benchmark results for four training-free generation pipelines with OPT-Kraken.}\label{tab:appendix-trainfree-detail}\\
\toprule
Benchmark & Initial & Final & Rescue & Regr. & Error Coverage \\
\midrule
\endfirsthead
\multicolumn{6}{l}{\tablename~\thetable\ (continued)}\\
\toprule
Benchmark & Initial & Final & Rescue & Regr. & Error Coverage \\
\midrule
\endhead
\midrule\multicolumn{6}{r}{Continued on next page}\\\endfoot
\bottomrule\endlastfoot
\multicolumn{6}{@{}l}{\textbf{Zero-shot}}\\
MAMO Complex & 25.1\% & 55.5\% & 40.5\% & 0.0\% & 84.2\% \\
IndustryOR & 50.6\% & 72.3\% & 43.9\% & 0.0\% & 100.0\% \\
ComplexOR & 11.8\% & 70.6\% & 66.7\% & 0.0\% & 100.0\% \\
OptMATH & 23.5\% & 48.8\% & 33.1\% & 0.0\% & 99.2\% \\
NLP4LP & 73.6\% & 80.6\% & 29.7\% & 1.1\% & 100.0\% \\
BWOR & 54.9\% & 67.1\% & 27.0\% & 0.0\% & 100.0\% \\
\textbf{Overall} & 44.8\% & 64.9\% & 36.9\% & 0.6\% & 94.1\% \\
\addlinespace[2pt]\midrule\addlinespace[1pt]
\multicolumn{6}{@{}l}{\textbf{ORMind}}\\
MAMO Complex & 49.3\% & 56.9\% & 15.9\% & 1.0\% & 97.4\% \\
IndustryOR & 60.2\% & 66.3\% & 15.2\% & 0.0\% & 100.0\% \\
ComplexOR & 52.9\% & 64.7\% & 25.0\% & 0.0\% & 100.0\% \\
OptMATH & 46.4\% & 50.0\% & 7.9\% & 1.3\% & 97.7\% \\
NLP4LP & 74.0\% & 79.3\% & 23.8\% & 1.1\% & 100.0\% \\
BWOR & 61.0\% & 72.0\% & 28.1\% & 0.0\% & 91.3\% \\
\textbf{Overall} & 58.6\% & 64.9\% & 16.6\% & 0.9\% & 97.8\% \\
\addlinespace[2pt]\midrule\addlinespace[1pt]
\multicolumn{6}{@{}l}{\textbf{OptiMUS-0.3}}\\
MAMO Complex & 39.8\% & 58.3\% & 30.7\% & 0.0\% & 97.4\% \\
IndustryOR & 48.2\% & 53.0\% & 14.0\% & 5.0\% & 95.2\% \\
ComplexOR & 47.1\% & 64.7\% & 33.3\% & 0.0\% & 83.3\% \\
OptMATH & 33.7\% & 39.2\% & 8.2\% & 0.0\% & 99.1\% \\
NLP4LP & 77.7\% & 81.4\% & 20.4\% & 1.1\% & 100.0\% \\
BWOR & 46.3\% & 53.7\% & 13.6\% & 0.0\% & 100.0\% \\
\textbf{Overall} & 51.7\% & 60.4\% & 19.1\% & 1.0\% & 98.0\% \\
\addlinespace[2pt]\midrule\addlinespace[1pt]
\multicolumn{6}{@{}l}{\textbf{SAC-Opt}}\\
MAMO Complex & 36.0\% & 54.0\% & 30.4\% & 3.9\% & 96.7\% \\
IndustryOR & 47.0\% & 57.8\% & 20.5\% & 0.0\% & 97.6\% \\
ComplexOR & 52.9\% & 64.7\% & 25.0\% & 0.0\% & 100.0\% \\
OptMATH & 21.7\% & 28.3\% & 9.2\% & 2.8\% & 99.2\% \\
NLP4LP & 79.3\% & 81.8\% & 12.0\% & 0.0\% & 100.0\% \\
BWOR & 40.2\% & 48.8\% & 14.3\% & 0.0\% & 97.8\% \\
\textbf{Overall} & 48.1\% & 57.2\% & 18.5\% & 1.0\% & 98.2\% \\
\end{longtable}

\endgroup

\subsection{Qwen3.8-27B Zero-Shot Generation}
\label{app:qwen-zero-shot-detail}
Table~\ref{tab:appendix-qwen-zero-shot-detail} complements the DeepSeek V4 Flash results in Appendix~\ref{app:trainfree-detail} with the complete six-benchmark breakdown for Qwen3.8-27B. Initial denotes the frozen zero-shot output; Final denotes the output retained by \method, including completed retries of abnormal exits. Qwen retains its recorded local-service configuration without an explicit thinking override in the chat template.
\begingroup
\fontsize{8.0pt}{8.8pt}\selectfont
\setlength{\tabcolsep}{3.4pt}
\setlength{\LTpre}{3pt}
\setlength{\LTpost}{3pt}
\begin{longtable}{@{}lrrrrr@{}}
\caption{Full six-benchmark results for Qwen3.8-27B zero-shot generation with OPT-Kraken.}\label{tab:appendix-qwen-zero-shot-detail}\\
\toprule
Benchmark & Initial & Final & Rescue & Regr. & Error Coverage \\
\midrule
\endfirsthead
\multicolumn{6}{l}{\tablename~\thetable\ (continued)}\\
\toprule
Benchmark & Initial & Final & Rescue & Regr. & Error Coverage \\
\midrule
\endhead
\midrule\multicolumn{6}{r}{Continued on next page}\\\endfoot
\bottomrule\endlastfoot
\multicolumn{6}{@{}l}{\textbf{Qwen3.8-27B Zero-shot + OPT-Kraken}}\\
MAMO Complex & 9.5\% & 23.2\% & 15.2\% & 0.0\% & 91.6\% \\
IndustryOR & 36.1\% & 45.8\% & 17.0\% & 3.3\% & 90.4\% \\
ComplexOR & 5.9\% & 41.2\% & 37.5\% & 0.0\% & 93.8\% \\
OptMATH & 11.4\% & 15.7\% & 4.8\% & 0.0\% & 97.3\% \\
NLP4LP & 45.5\% & 62.0\% & 30.3\% & 0.0\% & 98.4\% \\
BWOR & 42.7\% & 43.9\% & 4.3\% & 2.9\% & 73.3\% \\
\textbf{Overall} & 26.8\% & 38.2\% & 15.9\% & 0.9\% & 93.6\% \\
\end{longtable}

\endgroup

\subsection{Model Calls, Token Usage, and Runtime}
\label{app:resource-usage}
Table~\ref{tab:appendix-resource-usage} reports observed per-instance means from the saved DeepSeek V4 Flash invocation traces for each method's native workflow; it does not impose equal realized call counts. Differences arise from workflow structure and native stopping rules. Prompt and completion tokens are accumulated over every model request made for an instance. Full-prompt accounting counts all prompt tokens; the cache-aware view separates cached and uncached prompt tokens and reports the newly processed token volume as uncached prompt plus completion. Cached tokens remain part of API usage. Displayed token columns are independently rounded from the unrounded per-instance means, so displayed components can differ from a displayed total by one token.

Elapsed time is the end-to-end duration within each method's invocation boundary. Prompt-based Repair and CoT-based Repair each include one repair request, output processing, and candidate execution. Multi-stage Prompt-based Repair additionally includes generic diagnosis, independent review, review-triggered revision/re-review rounds, and final candidate execution. Because this baseline reused the saved first three stages and reran only initial review failures, its $11.90$-second end-to-end mean is reconstructed by combining the observed request time with the corresponding non-model overhead; model requests account for $10.48$ seconds on average. For \method, elapsed time covers post-generation detection, repair, execution, artifact audit, and retained-output selection. Across the 801 DeepSeek V4 Flash zero-shot artifacts, \method averages $15.70$ seconds per instance end to end (median $9.48$ seconds); recorded model-request time accounts for $15.21$ seconds on average.

\paragraph{Prompt-based repair baselines and inference budget.}
Table~\ref{tab:appendix-resource-usage} reports the three generic repair protocols under the same $(P/M_0/S_0)$, backbone, and decoding settings; Appendices~\ref{app:repair-prompt} and~\ref{app:enhanced-repair-prompts} give their prompts and call flows. Multi-stage averages $3.71$ logical calls, slightly more than \method's $3.65$. In contrast to these generic baselines, \method's calls implement evidence extraction, joint consistency checking, granularity selection, candidate generation, and artifact auditing. The table therefore exposes the computation profile of one-shot, explicit-CoT, generic multi-stage, and mechanism-structured repair; Table~\ref{tab:ablation} and the fixed-entry ablations in Appendix~\ref{app:sacopt-local} examine the three mechanism contributions.

\begingroup
\fontsize{7.7pt}{8.5pt}\selectfont
\setlength{\tabcolsep}{4.0pt}
\begin{classificationtable}
\centering
\caption{Observed DeepSeek V4 Flash call, token, and runtime profile. All values are means per instance over the saved formal invocation traces.}
\label{tab:appendix-resource-usage}
\textit{(a) Full-prompt accounting and end-to-end time.}\\[-2pt]
\begin{tabular}{@{}lrrrrr@{}}
\toprule
Method & Calls & Prompt & Completion & Total & Time (s) \\
\midrule
\multicolumn{6}{@{}l}{\textit{Generation workflows}} \\[-1pt]
Zero-shot & 2.01 & 943 & 1,281 & 2,224 & 9.01 \\
ORMind & 5.81 & 6,903 & 5,077 & 11,980 & 36.70 \\
OptiMUS-0.3 & 28.02 & 32,398 & 2,215 & 34,614 & 42.25 \\
SAC-Opt & 39.59 & 51,105 & 2,022 & 53,127 & 53.19 \\
\midrule
\multicolumn{6}{@{}l}{\textit{Repair methods}} \\[-1pt]
Prompt-based Repair & 1.00 & 1,644 & 1,103 & 2,747 & 7.15 \\
CoT-based Repair & 1.00 & 1,907 & 1,326 & 3,232 & 6.43 \\
Multi-stage Prompt-based Repair & 3.71 & 10,414 & 2,360 & 12,774 & 11.90 \\
\method & 3.65 & 31,089 & 2,911 & 34,000 & 15.70 \\
\bottomrule
\end{tabular}
\end{classificationtable}
\begin{classificationtable}
\centering
\textit{(b) Prompt-cache decomposition.}\\[-2pt]
\begin{tabular}{@{}lrrrrr@{}}
\toprule
Method & Prompt & Cached & Uncached & Completion & Uncached $+$ completion \\
\midrule
\multicolumn{6}{@{}l}{\textit{Generation workflows}} \\[-1pt]
Zero-shot & 943 & 60 & 883 & 1,281 & 2,163 \\
ORMind & 6,903 & 309 & 6,594 & 5,077 & 11,671 \\
OptiMUS-0.3 & 32,398 & 14,825 & 17,573 & 2,215 & 19,789 \\
SAC-Opt & 51,105 & 25,772 & 25,333 & 2,022 & 27,355 \\
\midrule
\multicolumn{6}{@{}l}{\textit{Repair methods}} \\[-1pt]
Prompt-based Repair & 1,644 & 102 & 1,542 & 1,103 & 2,646 \\
CoT-based Repair & 1,907 & 146 & 1,760 & 1,326 & 3,086 \\
Multi-stage Prompt-based Repair & 10,414 & 2,492 & 7,922 & 2,360 & 10,282 \\
\method & 31,089 & 23,215 & 7,874 & 2,911 & 10,785 \\
\bottomrule
\end{tabular}
\end{classificationtable}

\endgroup

\subsection{Mechanism-Wise and Fixed-Entry Ablations}
\label{app:sacopt-local}
Table~\ref{tab:appendix-mechanism-deltas} derives signed changes from Table~\ref{tab:ablation} and introduces no new experimental runs. Error Coverage is finalized before hierarchical repair; because B and C act downstream, both retain Full's 94.1\%, so their zero Coverage deltas are expected by design.

\begin{classificationtable}
  \centering
  \caption{Mechanism-wise ablation changes relative to Full (percentage points).}
  \label{tab:appendix-mechanism-deltas}
\small
\setlength{\tabcolsep}{5pt}
\begin{tabular}{@{}lrrrr@{}}
\toprule
Mechanism removed & Final Acc. $\Delta$ & Rescue $\Delta$ & \shortstack{Error\\Coverage $\Delta$} & Regr. $\Delta$ \\
\midrule
Dual-evidence diagnosis & $-14.7$ & $-24.5$ & $-7.0$ & $+2.7$ \\
Granularity selection & $-9.2$ & $-16.3$ & $0.0$ & $+0.5$ \\
Contracted local repair & $-7.1$ & $-12.5$ & $0.0$ & $+0.5$ \\
\bottomrule
\end{tabular}

\end{classificationtable}

\noindent\textbf{Fixed-entry local repair.}\par\smallskip
\begin{classificationtable}
  \centering
  \caption{Controlled fixed-entry ablation of contract-based local repair. Both variants use the same pipeline and maximum correction budget; Generic omits RepairCard, RepairTask, and Contract.}
  \label{tab:local}
\small
\setlength{\tabcolsep}{4pt}
\begin{tabular}{@{}lrrr@{}}
\toprule
Local-entry repairer & Rescued & Rescue (\%) & Regr. (\%) \\
\midrule
Evidence-constrained local repair & \textbf{11/40} & \textbf{27.5} & 0.0 \\
Generic repair without evidence & 3/40 & 7.5 & 0.0 \\
\bottomrule
\end{tabular}

\end{classificationtable}

With detection, routing, budget, and verification held fixed, contracts raise Rescue from 7.5\% to 27.5\% (Table~\ref{tab:local}).

The controlled ablation in Table~\ref{tab:local} uses identical frozen detection results, routing decisions, local-entry cases, and maximum repair attempts. Both variants retain the same execution feedback, artifact audit, safe-commit predicates, structural and final-direct branches, backbone, temperature, and decoding settings. The sole intervention is the local repair mechanism: the contract-based variant uses RepairCard, RepairTask, and Contract, whereas Generic uses an unconstrained local rewriting prompt under the same maximum correction budget. Early stopping remains enabled for both. Other repair paths partly compensate for local failures in the full-system result: the 20.0-point fixed-entry Rescue gap becomes a 12.5-point end-to-end Rescue gap between Full and Ablation C (36.9\% versus 24.4\%) because Ablation C changes only local-entry cases, Generic retains some repairs, structural or final-direct repair can recover local failures, and the final rate is computed over all 801 artifacts.

We additionally compare local correction on the same 140 frozen DeepSeek V4 Flash Plain Two-Stage artifacts: 40 erroneous artifacts for Rescue and 100 correct controls for Regression. Detection and routing are held fixed; every method begins from the same $P/M_0/S_0$ tuple and local entry. All methods use DeepSeek V4 Flash at temperature 0 with thinking disabled and are bounded by SAC-Opt's native maximum correction schedule: five repair rounds and three execution-debug calls. Native early stopping is preserved. Unsafe commits are retained edits that remain incorrect under the evaluator in Section~\ref{sec:settings}; Calls are per-artifact means over all 140 inputs and cover only processing after the fixed entry.

\begin{classificationtable}
  \centering
  \caption{Local repair under the same fixed entry and maximum correction budget. Rescued is reported over 40 erroneous artifacts and Regr. over 100 correct controls.}
  \label{tab:appendix-sacopt-local}
  \fontsize{7.7pt}{8.5pt}\selectfont
  \setlength{\tabcolsep}{3.3pt}
  \begin{tabular}{@{}lrrrrr@{}}
    \toprule
    Local repair mechanism & Rescued & Rescue & Regr. & Unsafe commits & Calls \\
    \midrule
    SAC-Opt semantic-anchor correction & 1/40 & 2.5\% & 2.0\% & 5/12 (41.7\%) & 1.58 \\
    \method\ evidence-constrained local repair & \textbf{11/40} & \textbf{27.5\%} & \textbf{0.0\%} & \textbf{5/47 (10.6\%)} & 5.08 \\
    \bottomrule
  \end{tabular}
\end{classificationtable}

\method\ rescues 11 times as many erroneous artifacts as SAC-Opt (11/40 versus 1/40) while maintaining zero Regression. The paired Rescue outcomes favor \method\ (11 Rescues exclusive to \method, one exclusive to SAC-Opt, and none shared; two-sided exact McNemar $p=0.00635$). SAC-Opt's lower mean call count primarily reflects coverage: its released local correction loop does not initiate repair for 123 of the 140 artifacts (87.9\%), whose fixed targets are outside constraint correction. On the 17 supported constraint targets, it rescues 1/8 erroneous artifacts and regresses 2/9 correct controls. Thus, the comparison provides additional evidence for the contribution of executable edit boundaries and candidate verification at the local repair stage.

\section{Hallucination Taxonomy and Error Profiles}
\label{app:hallucination-classification}
We expand the eight hallucination groups in Figure~\ref{fig:hallucination-distribution} into all 24 operational types. Each cell gives the number of pre-repair findings assigned to a frozen type, followed by its percentage of that column's total. An artifact may contribute more than one finding; percentages describe the classified findings, not the frequency of erroneous problems or an additional correctness evaluation.

\paragraph{Taxonomy and RepairCard construction.}
The expert process was used to develop the hallucination taxonomy and RepairCards, not to assign correctness labels to the formal evaluation outputs. Optimization-modeling experts, including two doctoral researchers, first reviewed naturally generated problem--model--code artifacts from multiple development benchmarks and generation models. They drafted categories around objectives, decision variables, constraints, data mapping, code implementation, and result extraction. Multiple LLMs, including DeepSeek, served as discovery aids for surfacing uncovered patterns, consolidating near-synonymous categories, and identifying boundary cases. We also created a small set of controlled elicitation cases by changing directions, indices, ratios, units, or constraint conditions to induce rare failure modes and stress-test coverage. The experts jointly refined category boundaries, inclusion and exclusion rules, positive and negative examples, and corresponding card eligibility conditions until reaching consensus. This was an iterative design process rather than independent annotation of a fixed evaluation cohort. Controlled cases were used only for taxonomy and card development; they are excluded from the reported hallucination frequencies and from all Rescue and Regression results.

\paragraph{Development--evaluation separation and freeze point.}
Every instance used to design the taxonomy, RepairCards, or routing rules came from development benchmarks that are disjoint from the six benchmarks used for formal evaluation. After the authors finalized the rule definitions and analyzed the separate development benchmarks, the O01--O24 taxonomy, complete RepairCard catalog, and routing rules were frozen before any formal evaluation on the six reported benchmarks. None of these components was subsequently revised in response to errors or outcomes observed on the evaluation benchmarks.

\paragraph{Use of the frozen taxonomy in formal evaluation.}
After the taxonomy was frozen, the framework assigned its pre-repair diagnostic findings to O01--O24 under the fixed category and card definitions; the tables below summarize these assignments. This analysis characterizes the forms of modeling hallucination encountered by the framework and does not rescore benchmark outputs. The experts did not provide Initial, Final, Rescue, or Regression labels. As specified in Section~\ref{sec:settings}, those main-table outcomes are determined solely by the required objective value or solve status and the stated tolerances. Taxonomy construction and benchmark scoring therefore remain separate; these tables characterize taxonomy coverage and error composition rather than a second set of benchmark correctness labels.

\subsection{DeepSeek V4 Flash across Eight Upstreams}
\label{app:hallucination-deepseek}
Tables~\ref{tab:appendix-hallucination-trainfree} and~\ref{tab:appendix-hallucination-finetuned} retain the four training-free and four fine-tuned upstreams separately. Their totals are 688 and 950 findings, respectively, matching Figure~\ref{fig:hallucination-distribution}. Model abbreviations follow Table~\ref{tab:finetuned}.

\begingroup
  \fontsize{6.8pt}{7.5pt}\selectfont
  \setlength{\tabcolsep}{2pt}
  \setlength{\LTpre}{5pt}
  \setlength{\LTpost}{5pt}
\begin{longtable}{@{}p{0.045\linewidth}p{0.370\linewidth}*{4}{>{\raggedleft\arraybackslash}p{0.127\linewidth}}@{}}
\caption{O01--O24 pre-repair error profile: DeepSeek V4 Flash with training-free upstreams.}\label{tab:appendix-hallucination-trainfree} \\
\toprule
ID & Hallucination type & ORMind & OptiMUS & SAC-Opt & Zero-shot \\
\midrule
\endfirsthead
\multicolumn{6}{@{}l}{\tablename~\thetable\ (continued)} \\
\toprule
ID & Hallucination type & ORMind & OptiMUS & SAC-Opt & Zero-shot \\
\midrule
\endhead
\midrule
\multicolumn{6}{r}{Continued on next page} \\
\endfoot
\bottomrule
\endlastfoot
O01 & Objective goal or sense & 5 (5.9) & 13 (6.2) & 13 (7.1) & 2 (0.9) \\
O02 & Objective terms or coefficients & 6 (7.1) & 19 (9.0) & 21 (11.5) & 6 (2.8) \\
O03 & Risk or multiobjective structure & 1 (1.2) & 3 (1.4) & 3 (1.6) & 2 (0.9) \\
O04 & Hard/soft requirement role & 0 (0.0) & 0 (0.0) & 1 (0.5) & 0 (0.0) \\
\addlinespace[2pt]
O05 & Decision objects or indices & 1 (1.2) & 2 (1.0) & 0 (0.0) & 0 (0.0) \\
O06 & Continuous-to-discrete domain & 0 (0.0) & 0 (0.0) & 0 (0.0) & 2 (0.9) \\
O07 & Discrete-to-continuous domain & 2 (2.4) & 0 (0.0) & 0 (0.0) & 1 (0.5) \\
O08 & Variable bounds or allowed sets & 3 (3.5) & 25 (11.9) & 8 (4.4) & 3 (1.4) \\
\addlinespace[2pt]
O09 & Missing required constraints & 5 (5.9) & 23 (11.0) & 12 (6.6) & 2 (0.9) \\
O10 & Unsupported constraints & 2 (2.4) & 7 (3.3) & 3 (1.6) & 4 (1.9) \\
\addlinespace[2pt]
O11 & Equality or conservation & 12 (14.1) & 27 (12.9) & 33 (18.1) & 7 (3.3) \\
O12 & Scalar bounds or strictness & 9 (10.6) & 12 (5.7) & 15 (8.2) & 12 (5.7) \\
\addlinespace[2pt]
O13 & Pairwise comparative relations & 0 (0.0) & 3 (1.4) & 0 (0.0) & 1 (0.5) \\
O14 & Share of a combined total & 2 (2.4) & 0 (0.0) & 0 (0.0) & 2 (0.9) \\
O15 & Ratio to another quantity & 0 (0.0) & 2 (1.0) & 0 (0.0) & 2 (0.9) \\
O16 & Row algebra or role binding & 7 (8.2) & 20 (9.5) & 19 (10.4) & 8 (3.8) \\
\addlinespace[2pt]
O17 & Index or quantifier scope & 13 (15.3) & 29 (13.8) & 13 (7.1) & 25 (11.8) \\
O18 & Logic or activation & 3 (3.5) & 7 (3.3) & 3 (1.6) & 0 (0.0) \\
O19 & Network or temporal structure & 0 (0.0) & 11 (5.2) & 18 (9.9) & 0 (0.0) \\
\addlinespace[2pt]
O20 & Parameter grounding & 1 (1.2) & 1 (0.5) & 0 (0.0) & 0 (0.0) \\
O21 & Units or time scope & 1 (1.2) & 2 (1.0) & 0 (0.0) & 1 (0.5) \\
\addlinespace[2pt]
O22 & Model-to-code construction & 10 (11.8) & 4 (1.9) & 8 (4.4) & 112 (53.1) \\
O23 & Code index/set materialization & 0 (0.0) & 0 (0.0) & 12 (6.6) & 10 (4.7) \\
O24 & Solve, extraction or reporting & 2 (2.4) & 0 (0.0) & 0 (0.0) & 9 (4.3) \\
\midrule
\multicolumn{2}{@{}l}{Total findings} & 85 (100.0) & 210 (100.0) & 182 (100.0) & 211 (100.0) \\
\end{longtable}

\endgroup

\begingroup
  \fontsize{6.8pt}{7.5pt}\selectfont
  \setlength{\tabcolsep}{2pt}
  \setlength{\LTpre}{5pt}
  \setlength{\LTpost}{5pt}
\begin{longtable}{@{}p{0.045\linewidth}p{0.370\linewidth}*{4}{>{\raggedleft\arraybackslash}p{0.127\linewidth}}@{}}
\caption{O01--O24 pre-repair error profile: DeepSeek V4 Flash with fine-tuned upstreams.}\label{tab:appendix-hallucination-finetuned} \\
\toprule
ID & Hallucination type & SIRL & LLMOPT & ORLM & OptMATH \\
\midrule
\endfirsthead
\multicolumn{6}{@{}l}{\tablename~\thetable\ (continued)} \\
\toprule
ID & Hallucination type & SIRL & LLMOPT & ORLM & OptMATH \\
\midrule
\endhead
\midrule
\multicolumn{6}{r}{Continued on next page} \\
\endfoot
\bottomrule
\endlastfoot
O01 & Objective goal or sense & 3 (1.8) & 8 (2.9) & 18 (6.4) & 12 (5.3) \\
O02 & Objective terms or coefficients & 13 (7.9) & 25 (9.1) & 30 (10.6) & 31 (13.6) \\
O03 & Risk or multiobjective structure & 0 (0.0) & 3 (1.1) & 0 (0.0) & 3 (1.3) \\
O04 & Hard/soft requirement role & 0 (0.0) & 0 (0.0) & 0 (0.0) & 0 (0.0) \\
\addlinespace[2pt]
O05 & Decision objects or indices & 1 (0.6) & 1 (0.4) & 3 (1.1) & 5 (2.2) \\
O06 & Continuous-to-discrete domain & 1 (0.6) & 1 (0.4) & 0 (0.0) & 1 (0.4) \\
O07 & Discrete-to-continuous domain & 0 (0.0) & 0 (0.0) & 0 (0.0) & 1 (0.4) \\
O08 & Variable bounds or allowed sets & 1 (0.6) & 12 (4.3) & 1 (0.4) & 3 (1.3) \\
\addlinespace[2pt]
O09 & Missing required constraints & 29 (17.7) & 29 (10.5) & 30 (10.6) & 27 (11.8) \\
O10 & Unsupported constraints & 7 (4.3) & 7 (2.5) & 3 (1.1) & 0 (0.0) \\
\addlinespace[2pt]
O11 & Equality or conservation & 48 (29.3) & 43 (15.6) & 85 (30.1) & 58 (25.4) \\
O12 & Scalar bounds or strictness & 12 (7.3) & 17 (6.2) & 17 (6.0) & 17 (7.5) \\
\addlinespace[2pt]
O13 & Pairwise comparative relations & 2 (1.2) & 4 (1.4) & 3 (1.1) & 1 (0.4) \\
O14 & Share of a combined total & 0 (0.0) & 0 (0.0) & 0 (0.0) & 0 (0.0) \\
O15 & Ratio to another quantity & 3 (1.8) & 4 (1.4) & 1 (0.4) & 1 (0.4) \\
O16 & Row algebra or role binding & 24 (14.6) & 34 (12.3) & 33 (11.7) & 32 (14.0) \\
\addlinespace[2pt]
O17 & Index or quantifier scope & 7 (4.3) & 30 (10.9) & 30 (10.6) & 16 (7.0) \\
O18 & Logic or activation & 2 (1.2) & 3 (1.1) & 5 (1.8) & 0 (0.0) \\
O19 & Network or temporal structure & 8 (4.9) & 13 (4.7) & 6 (2.1) & 6 (2.6) \\
\addlinespace[2pt]
O20 & Parameter grounding & 0 (0.0) & 0 (0.0) & 0 (0.0) & 0 (0.0) \\
O21 & Units or time scope & 1 (0.6) & 0 (0.0) & 1 (0.4) & 2 (0.9) \\
\addlinespace[2pt]
O22 & Model-to-code construction & 2 (1.2) & 38 (13.8) & 14 (5.0) & 9 (3.9) \\
O23 & Code index/set materialization & 0 (0.0) & 4 (1.4) & 2 (0.7) & 3 (1.3) \\
O24 & Solve, extraction or reporting & 0 (0.0) & 0 (0.0) & 0 (0.0) & 0 (0.0) \\
\midrule
\multicolumn{2}{@{}l}{Total findings} & 164 (100.0) & 276 (100.0) & 282 (100.0) & 228 (100.0) \\
\end{longtable}

\endgroup

\subsection{Qwen and GPT Zero-Shot Outputs}
\label{app:hallucination-backbones}
Table~\ref{tab:appendix-hallucination-backbones} classifies the frozen pre-repair Zero-shot outputs. Qwen contributes 319 classified findings from 302 artifacts. The GPT column uses the existing 339-case review of unrecovered outputs: 159 artifacts have 200 retained O01--O24 findings. The additional Qwen Rescue was already classified as O24; recovering it changes the repair outcome, not the count of pre-repair findings. These are within-column classification profiles over the stated coverage, rather than a matched-cohort comparison of the two backbones.

\begingroup
  \fontsize{7.4pt}{8.3pt}\selectfont
  \setlength{\tabcolsep}{3pt}
  \setlength{\LTpre}{5pt}
  \setlength{\LTpost}{5pt}
\begin{longtable}{@{}p{0.05\linewidth}p{0.50\linewidth}*{2}{>{\raggedleft\arraybackslash}p{0.18\linewidth}}@{}}
\caption{O01--O24 pre-repair error profile: Qwen3.8-27B and GPT-5.6 Sol Zero-shot outputs.}\label{tab:appendix-hallucination-backbones} \\
\toprule
ID & Hallucination type & Qwen3.8-27B & GPT-5.6 Sol \\
\midrule
\endfirsthead
\multicolumn{4}{@{}l}{\tablename~\thetable\ (continued)} \\
\toprule
ID & Hallucination type & Qwen3.8-27B & GPT-5.6 Sol \\
\midrule
\endhead
\midrule
\multicolumn{4}{r}{Continued on next page} \\
\endfoot
\bottomrule
\endlastfoot
O01 & Objective goal or sense & 1 (0.3) & 1 (0.5) \\
O02 & Objective terms or coefficients & 2 (0.6) & 24 (12.0) \\
O03 & Risk or multiobjective structure & 3 (0.9) & 3 (1.5) \\
O04 & Hard/soft requirement role & 0 (0.0) & 4 (2.0) \\
\addlinespace[2pt]
O05 & Decision objects or indices & 0 (0.0) & 1 (0.5) \\
O06 & Continuous-to-discrete domain & 4 (1.3) & 0 (0.0) \\
O07 & Discrete-to-continuous domain & 0 (0.0) & 0 (0.0) \\
O08 & Variable bounds or allowed sets & 0 (0.0) & 4 (2.0) \\
\addlinespace[2pt]
O09 & Missing required constraints & 0 (0.0) & 4 (2.0) \\
O10 & Unsupported constraints & 1 (0.3) & 5 (2.5) \\
\addlinespace[2pt]
O11 & Equality or conservation & 3 (0.9) & 13 (6.5) \\
O12 & Scalar bounds or strictness & 7 (2.2) & 5 (2.5) \\
\addlinespace[2pt]
O13 & Pairwise comparative relations & 1 (0.3) & 0 (0.0) \\
O14 & Share of a combined total & 1 (0.3) & 1 (0.5) \\
O15 & Ratio to another quantity & 1 (0.3) & 0 (0.0) \\
O16 & Row algebra or role binding & 6 (1.9) & 8 (4.0) \\
\addlinespace[2pt]
O17 & Index or quantifier scope & 33 (10.3) & 8 (4.0) \\
O18 & Logic or activation & 0 (0.0) & 9 (4.5) \\
O19 & Network or temporal structure & 0 (0.0) & 2 (1.0) \\
\addlinespace[2pt]
O20 & Parameter grounding & 0 (0.0) & 9 (4.5) \\
O21 & Units or time scope & 0 (0.0) & 2 (1.0) \\
\addlinespace[2pt]
O22 & Model-to-code construction & 25 (7.8) & 39 (19.5) \\
O23 & Code index/set materialization & 11 (3.4) & 6 (3.0) \\
O24 & Solve, extraction or reporting & 220 (69.0) & 52 (26.0) \\
\midrule
\multicolumn{2}{@{}l}{Total findings} & 319 (100.0) & 200 (100.0) \\
\end{longtable}

\endgroup

\subsection{Outcome-Conditioned O01--O24 Composition}
\label{app:hallucination-outcomes}
Tables~\ref{tab:appendix-outcome-qwen}--\ref{tab:appendix-outcome-gpt} disaggregate the eight groups in Figure~\ref{fig:hallucination-distribution} into all 24 operational types and condition them on the final outcome for each Zero-shot backbone. The Rescue column is the within-outcome distribution of retained pre-repair operational findings associated with artifacts that became correct. The Regression column instead describes the diagnostic types that triggered harmful edits to originally correct artifacts; it characterizes the composition of these false-alarm triggers and does not imply that the original artifacts contained true hallucinations. Each column is normalized over its retained classified findings. Evidence that remained unclassified or outside the O01--O24 boundary is excluded, and an artifact may contribute multiple independent findings. Values are percentages only and are rounded independently, so a displayed column may differ from 100\% by 0.1 percentage point. The tables do not repeat overall Rescue, Regression, or Error Coverage rates.

\begingroup
  \fontsize{6.5pt}{7.1pt}\selectfont
  \setlength{\tabcolsep}{5pt}
  \setlength{\LTpre}{5pt}
  \setlength{\LTpost}{5pt}
\begin{longtable}{@{}p{0.05\linewidth}p{0.50\linewidth}>{\raggedleft\arraybackslash}p{0.17\linewidth}>{\raggedleft\arraybackslash}p{0.18\linewidth}@{}}
\caption{O01--O24 within-outcome hallucination composition for Qwen3.8-27B. Only percentages of retained classified findings are shown.}\label{tab:appendix-outcome-qwen} \\
\toprule
ID & Hallucination type & Rescue (\%) & Regression (\%) \\
\midrule
\endfirsthead
\multicolumn{4}{@{}l}{\tablename~\thetable\ (continued)} \\
\toprule
ID & Hallucination type & Rescue (\%) & Regression (\%) \\
\midrule
\endhead
\midrule
\multicolumn{4}{r}{Continued on next page} \\
\endfoot
\bottomrule
\endlastfoot
O01 & Objective goal or sense & 1.3\% & 0.0\% \\
O02 & Objective terms or coefficients & 0.0\% & 0.0\% \\
O03 & Risk or multiobjective structure & 0.0\% & 0.0\% \\
O04 & Hard/soft requirement role & 0.0\% & 0.0\% \\
\addlinespace[2pt]
O05 & Decision objects or indices & 0.0\% & 0.0\% \\
O06 & Continuous-to-discrete domain & 0.0\% & 0.0\% \\
O07 & Discrete-to-continuous domain & 0.0\% & 0.0\% \\
O08 & Variable bounds or allowed sets & 0.0\% & 0.0\% \\
\addlinespace[2pt]
O09 & Missing required constraints & 0.0\% & 0.0\% \\
O10 & Unsupported constraints & 0.0\% & 0.0\% \\
\addlinespace[2pt]
O11 & Equality or conservation & 0.0\% & 50.0\% \\
O12 & Scalar bounds or strictness & 0.0\% & 50.0\% \\
\addlinespace[2pt]
O13 & Pairwise comparative relations & 0.0\% & 0.0\% \\
O14 & Share of a combined total & 0.0\% & 0.0\% \\
O15 & Ratio to another quantity & 0.0\% & 0.0\% \\
O16 & Row algebra or role binding & 2.5\% & 0.0\% \\
\addlinespace[2pt]
O17 & Index or quantifier scope & 3.8\% & 0.0\% \\
O18 & Logic or activation & 0.0\% & 0.0\% \\
O19 & Network or temporal structure & 0.0\% & 0.0\% \\
\addlinespace[2pt]
O20 & Parameter grounding & 0.0\% & 0.0\% \\
O21 & Units or time scope & 0.0\% & 0.0\% \\
\addlinespace[2pt]
O22 & Model-to-code construction & 3.8\% & 0.0\% \\
O23 & Code index/set materialization & 2.5\% & 0.0\% \\
O24 & Solve, extraction or reporting & 86.1\% & 0.0\% \\
\end{longtable}

\endgroup

\begingroup
  \fontsize{6.5pt}{7.1pt}\selectfont
  \setlength{\tabcolsep}{5pt}
  \setlength{\LTpre}{5pt}
  \setlength{\LTpost}{5pt}
\begin{longtable}{@{}p{0.05\linewidth}p{0.50\linewidth}>{\raggedleft\arraybackslash}p{0.17\linewidth}>{\raggedleft\arraybackslash}p{0.18\linewidth}@{}}
\caption{O01--O24 within-outcome hallucination composition for DeepSeek V4 Flash. Only percentages of retained classified findings are shown.}\label{tab:appendix-outcome-deepseek} \\
\toprule
ID & Hallucination type & Rescue (\%) & Regression (\%) \\
\midrule
\endfirsthead
\multicolumn{4}{@{}l}{\tablename~\thetable\ (continued)} \\
\toprule
ID & Hallucination type & Rescue (\%) & Regression (\%) \\
\midrule
\endhead
\midrule
\multicolumn{4}{r}{Continued on next page} \\
\endfoot
\bottomrule
\endlastfoot
O01 & Objective goal or sense & 0.0\% & 0.0\% \\
O02 & Objective terms or coefficients & 0.9\% & 0.0\% \\
O03 & Risk or multiobjective structure & 0.0\% & 0.0\% \\
O04 & Hard/soft requirement role & 0.0\% & 0.0\% \\
\addlinespace[2pt]
O05 & Decision objects or indices & 0.0\% & 0.0\% \\
O06 & Continuous-to-discrete domain & 0.0\% & 100.0\% \\
O07 & Discrete-to-continuous domain & 0.0\% & 0.0\% \\
O08 & Variable bounds or allowed sets & 1.8\% & 0.0\% \\
\addlinespace[2pt]
O09 & Missing required constraints & 1.8\% & 0.0\% \\
O10 & Unsupported constraints & 0.9\% & 0.0\% \\
\addlinespace[2pt]
O11 & Equality or conservation & 3.6\% & 0.0\% \\
O12 & Scalar bounds or strictness & 4.5\% & 0.0\% \\
\addlinespace[2pt]
O13 & Pairwise comparative relations & 0.0\% & 0.0\% \\
O14 & Share of a combined total & 0.0\% & 0.0\% \\
O15 & Ratio to another quantity & 0.0\% & 0.0\% \\
O16 & Row algebra or role binding & 0.9\% & 0.0\% \\
\addlinespace[2pt]
O17 & Index or quantifier scope & 11.8\% & 0.0\% \\
O18 & Logic or activation & 0.0\% & 0.0\% \\
O19 & Network or temporal structure & 0.0\% & 0.0\% \\
\addlinespace[2pt]
O20 & Parameter grounding & 0.0\% & 0.0\% \\
O21 & Units or time scope & 0.0\% & 0.0\% \\
\addlinespace[2pt]
O22 & Model-to-code construction & 66.4\% & 0.0\% \\
O23 & Code index/set materialization & 5.5\% & 0.0\% \\
O24 & Solve, extraction or reporting & 1.8\% & 0.0\% \\
\end{longtable}

\endgroup

\begingroup
  \fontsize{6.5pt}{7.1pt}\selectfont
  \setlength{\tabcolsep}{5pt}
  \setlength{\LTpre}{5pt}
  \setlength{\LTpost}{5pt}
\begin{longtable}{@{}p{0.05\linewidth}p{0.50\linewidth}>{\raggedleft\arraybackslash}p{0.17\linewidth}>{\raggedleft\arraybackslash}p{0.18\linewidth}@{}}
\caption{O01--O24 within-outcome hallucination composition for GPT-5.6 Sol. Only percentages of retained classified findings are shown.}\label{tab:appendix-outcome-gpt} \\
\toprule
ID & Hallucination type & Rescue (\%) & Regression (\%) \\
\midrule
\endfirsthead
\multicolumn{4}{@{}l}{\tablename~\thetable\ (continued)} \\
\toprule
ID & Hallucination type & Rescue (\%) & Regression (\%) \\
\midrule
\endhead
\midrule
\multicolumn{4}{r}{Continued on next page} \\
\endfoot
\bottomrule
\endlastfoot
O01 & Objective goal or sense & 0.0\% & \textemdash \\
O02 & Objective terms or coefficients & 4.3\% & \textemdash \\
O03 & Risk or multiobjective structure & 0.0\% & \textemdash \\
O04 & Hard/soft requirement role & 0.0\% & \textemdash \\
\addlinespace[2pt]
O05 & Decision objects or indices & 0.0\% & \textemdash \\
O06 & Continuous-to-discrete domain & 0.0\% & \textemdash \\
O07 & Discrete-to-continuous domain & 0.0\% & \textemdash \\
O08 & Variable bounds or allowed sets & 0.0\% & \textemdash \\
\addlinespace[2pt]
O09 & Missing required constraints & 0.0\% & \textemdash \\
O10 & Unsupported constraints & 2.1\% & \textemdash \\
\addlinespace[2pt]
O11 & Equality or conservation & 8.5\% & \textemdash \\
O12 & Scalar bounds or strictness & 0.0\% & \textemdash \\
\addlinespace[2pt]
O13 & Pairwise comparative relations & 2.1\% & \textemdash \\
O14 & Share of a combined total & 2.1\% & \textemdash \\
O15 & Ratio to another quantity & 2.1\% & \textemdash \\
O16 & Row algebra or role binding & 8.5\% & \textemdash \\
\addlinespace[2pt]
O17 & Index or quantifier scope & 4.3\% & \textemdash \\
O18 & Logic or activation & 0.0\% & \textemdash \\
O19 & Network or temporal structure & 0.0\% & \textemdash \\
\addlinespace[2pt]
O20 & Parameter grounding & 0.0\% & \textemdash \\
O21 & Units or time scope & 0.0\% & \textemdash \\
\addlinespace[2pt]
O22 & Model-to-code construction & 12.8\% & \textemdash \\
O23 & Code index/set materialization & 0.0\% & \textemdash \\
O24 & Solve, extraction or reporting & 53.2\% & \textemdash \\
\end{longtable}

\endgroup

For GPT-5.6 Sol, Regression is zero and there are no samples; its Regression distribution is therefore undefined and is shown with dashes.

\section{End-to-End Repair Output}
\label{app:repair-output}
We return to the formal \texttt{industryor:prob\_085} trace introduced in Appendix~\ref{app:formal-artifact}. The following records show the detector output, the committed mathematical model, synchronized solver code, and the final audit result.

\subsection{Hallucination Record and Candidate Patch}
\label{app:detected-output}
The detector anchors the diagnosis to the price, raw-material cost, and variable-cost statements. It authorizes exactly one model-level target.

\artifactheading{Detected Hallucination and Minimum Complete Write Set}
\begin{lstlisting}[style=artifactcompact]
{
  "diagnosis": "M0's own source-derived objective arithmetic is inconsistent with its simplified coefficients (x1: stated 1.15, arithmetic gives 0.65, x2: stated 0.65, arithmetic gives 0.45).",
  "minimum_complete_write_set": [
    "objective"
  ],
  "model_refs": [
    "objective"
  ],
  "route": "M_SIDE_LOCAL",
  "solver_refs": [],
  "source_quotes": [
    "Healthy Pet Foods Company produces two types of dog food: Meaties and Yummies. Each pack of Meaties contains 2 pounds of grains and 3 pounds of meat; each pack of Yummies contains 3 pounds of grains and 1.5 pounds of meat. The company believes it can sell any quantity of dog food that it can produce. Meaties sell for $2.80 per pack, and Yummies sell for $2.00 per pack. The company's production is subject to several constraints. First, a maximum of 400,000 pounds of grains can be purchased each month at a price of $0.20 per pound of grains. A maximum of 300,000 pounds of meat can be purchased each month at a price of $0.50 per pound of meat. Additionally, a special machine is required to produce Meaties, with a monthly capacity of 90,000 packs. The variable costs for mixing and packaging dog food are $0.25 per pack (Meaties) and $0.20 per pack (Yummies). Detailed information is provided in Table B-1.",
    "|                    | Meaties      | Yummies    |",
    "| Price per pack     | $2.80        | $2.00      |",
    "| - Grains           | 2.0 lbs      | 3.0 lbs    |",
    "| - Meat             | 3.0 lbs      | 1.5 lbs    |",
    "| Variable cost      | $0.25/pack   | $0.20/pack |",
    "| Meaties capacity   | 90,000 packs/month |       |",
    "| Monthly available grains | 400,000 lbs |      |"
  ]
}
\end{lstlisting}
\artifactheading{Candidate Selection and Applied References}
\begin{lstlisting}[style=artifactcompact]
{
  "record_id": "industryor:prob_085",
  "detection_state": "ERROR",
  "selected_branch": "local_model_patch",
  "patch_record": {
    "allowed_refs": [
      "objective"
    ],
    "applied_refs": [
      "objective"
    ]
  },
  "candidate_accepted": true
}
\end{lstlisting}

\subsection{Final Mathematical Model and Solver Code}
\label{app:final-output}
Let $x_1$ and $x_2$ denote monthly packs of Meaties and Yummies. The committed model is
\[
\begin{aligned}
\max_{x_1,x_2\geq0}\quad &0.65x_1+0.45x_2\\
\text{s.t.}\quad &2x_1+3x_2\leq400{,}000,\\
&3x_1+1.5x_2\leq300{,}000,\\
&x_1\leq90{,}000.
\end{aligned}
\]
The ModelIR record below is the committed $M^*$. Its resource constraints, bounds, and variable domains are unchanged; only the objective coefficients differ from $M_0$.

\artifactheading{Committed Mathematical Model $M^*$ (ModelIR)}
\begin{lstlisting}[style=artifactcompact]
{
  "answer_rounding": {
    "digits": 8,
    "mode": "none"
  },
  "constraints": [
    {
      "expression": "((2.0)*x1 + (3.0)*x2) <= (400000.0)",
      "name": "c_0001_grains"
    },
    {
      "expression": "((3.0)*x1 + (1.5)*x2) <= (300000.0)",
      "name": "c_0002_meat"
    },
    {
      "expression": "((1.0)*x1) <= (90000.0)",
      "name": "c_0003_capacity"
    }
  ],
  "objective": {
    "expression": "(0.65)*x1 + (0.45)*x2"
  },
  "sense": "max",
  "variables": [
    {
      "lb": 0.0,
      "name": "x1",
      "type": "continuous",
      "ub": null
    },
    {
      "lb": 0.0,
      "name": "x2",
      "type": "continuous",
      "ub": null
    }
  ]
}
\end{lstlisting}

The synchronized $S^*$ below implements the Contract-approved model patch. Its public no-argument \texttt{solve\_model()} wrapper executes the committed ModelIR through \texttt{solve\_quantified\_model} and returns the same four-field record required from generated solver code.

\artifactheading{OPT-Kraken Solver Code $S^*$ with Public Interface Wrapper}
\begin{lstlisting}[style=artifactcompact,language=Python]
import json
from optkraken.quantified_model import solve_quantified_model

QUANTIFIED_MODEL = json.loads('{"answer_rounding": {"digits": 8, "mode": "none"}, "constraints": [{"expression": "((2.0)*x1 + (3.0)*x2) <= (400000.0)", "forall": [], "name": "c_0001_grains"}, {"expression": "((3.0)*x1 + (1.5)*x2) <= (300000.0)", "forall": [], "name": "c_0002_meat"}, {"expression": "((1.0)*x1) <= (90000.0)", "forall": [], "name": "c_0003_capacity"}], "objective": {"expression": "(0.65)*x1 + (0.45)*x2"}, "parameters": {}, "sense": "max", "sets": {"_ONE": [0]}, "variables": [{"indices": [], "lb": 0.0, "name": "x1", "type": "continuous", "ub": null}, {"indices": [], "lb": 0.0, "name": "x2", "type": "continuous", "ub": null}]}')
def solve_model():
    result = solve_quantified_model(QUANTIFIED_MODEL, 300.0, 0.0001)
    return {'status': result.status,
            'objective_value': result.objective_value,
            'reported_value': result.reported_value,
            'mip_gap': result.mip_gap}


if __name__ == '__main__':
    print(json.dumps(solve_model(), sort_keys=True))
\end{lstlisting}

\subsection{Execution, Artifact Audit, and Commit}
\label{app:audit-output}
The repaired program executes to optimality with objective $77{,}500$. The artifact audit returns \textsc{Pass}, and the candidate becomes the retained output.

\artifactheading{Execution Envelope and Artifact Audit}
\begin{lstlisting}[style=artifactcompact]
{
  "execution": {
    "execution_adapter": "in_process_modelir",
    "execution_success": true,
    "message": "Gurobi optimal",
    "mip_gap": null,
    "output_solution": "77500.0",
    "status": "optimal"
  },
  "semantic_audit": {
    "affirmative_non_errors_removed": [],
    "effect_findings_retracted": [],
    "findings": [],
    "self_retracted_findings": [],
    "source_quote_normalizations": [],
    "status": "PASS",
    "ungrounded_findings": [],
    "weak_domain_downgrades": []
  },
  "candidate_accepted": true
}
\end{lstlisting}

\end{document}